\documentclass[11pt,a4paper]{article}

\usepackage[truedimen,margin=32mm]{geometry} 

\usepackage{mathrsfs}
\usepackage{amssymb}
\usepackage{amsmath}
\usepackage{ascmac}
\usepackage{amsthm}
\usepackage[pdftex]{graphicx}
\usepackage{natbib}
\usepackage{setspace}
\usepackage{url}
\usepackage{here}
\usepackage{comment,booktabs}
\usepackage{lscape}
\usepackage{multirow}
\usepackage{algorithm, algorithmicx, algpseudocode}

\usepackage{titlesec}
\titleformat*{\section}{\large\bfseries}
\titleformat*{\subsection}{\it}

\theoremstyle{plain}

\theoremstyle{definition}

\newcommand{\diag}{\mathrm{diag}}

\title{{\bf Locally Adaptive Spatial Quantile Smoothing: Application to Monitoring Crime Density in Tokyo}}
\date{}

\begin{document}

\maketitle
\doublespacing

\vspace{-1.5cm}
\begin{center}
{\large Takahiro Onizuka$^{1}$, Shintaro Hashimoto$^{1}$ and Shonosuke Sugasawa$^{2}$}
\end{center}

\medskip
\noindent
$^{1}$Department of Mathematics, Hiroshima University\\
$^{2}$Faculty of Economics, Keio University

\medskip
\medskip
\medskip
\begin{center}
{\bf \large Abstract}
\end{center}

Spatial trend estimation under potential heterogeneity is an important problem to extract spatial characteristics and hazards such as criminal activity. By focusing on quantiles, which provide substantial information on distributions compared with commonly used summary statistics such as means, it is often useful to estimate not only the average trend but also the high (low) risk trend additionally. 
In this paper, we propose a Bayesian quantile trend filtering method to estimate the non-stationary trend of quantiles on graphs and apply it to crime data in Tokyo between 2013 and 2017. By modeling multiple observation cases, we can estimate the potential heterogeneity of spatial crime trends over multiple years in the application. 
To induce locally adaptive Bayesian inference on trends, we introduce general shrinkage priors for graph differences. Introducing so-called shadow priors with multivariate distribution for local scale parameters and mixture representation of the asymmetric Laplace distribution, we provide a simple Gibbs sampling algorithm to generate posterior samples. 
The numerical performance of the proposed method is demonstrated through simulation studies. 


\newpage

\section{Introduction}
\label{sec:1}

\begin{figure}[thbp]
\begin{center}
\includegraphics[width=\linewidth]{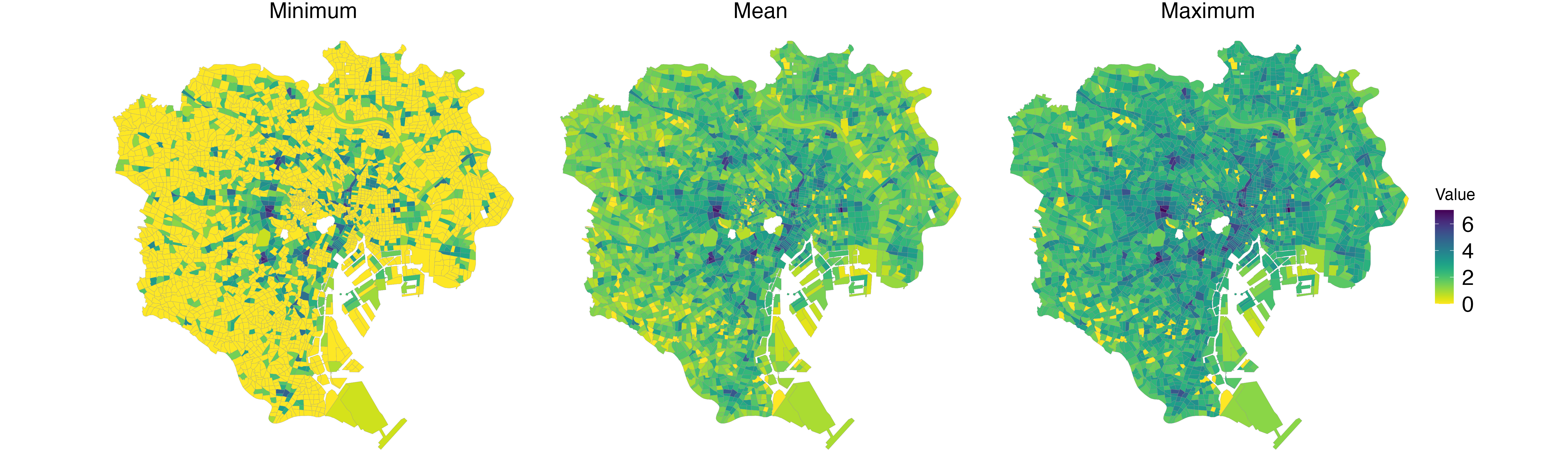}
\caption{Spatial plot of $\log(1+Y)$ for crime density $Y$. From left to right, the minimum, mean, and maximum values for each area over five years from 2013 to 2017.}
\label{tokyo_crime_minmeanmaxplot}
\end{center}
\end{figure}

Estimating the spatial trend of the number of crimes is vital to ensure community safety and to respond quickly to incidents. For example, more police may be assigned to areas with a lot of crimes than to areas with few crimes. Tokyo metropolitan police department mentioned that crime predictions have some effects: 1) Efficient development of police officers, 2) Realization of improved public safety, 3) Improving police operations efficiency, and 4) Conducting effective patrols. Inference on the crime risk for each area is an important task for crime data analysis, and it has been revealed that crime can be controlled more effectively and efficiently by concentrating police enforcement efforts on high-risk spots and time \citep[e.g.][]{braga2001effects}.
Since the number of crimes is often heterogeneous per region, the use of statistical models that take into account such heterogeneity is necessary. In Japan, University of Tsukuba Division of Policy and Planning Sciences Commons provides ``GIS database of several police-recorded crimes at O-aza, chome in Tokyo, 2009--2017". The data contain the number of various crimes from 2009 to 2017 as well as spatial information and the area for each region. Recently, \cite{hamura2021robust} and \cite{yano2021minimax} dealt with the data as zero-inflated count data and they proposed hierarchical Poisson models. It is known that crime data have spatial heterogeneity in the sense that most of the areas have little or no crime throughout multiple years, while others have a lot of crime yearly. 
Figure \ref{tokyo_crime_minmeanmaxplot} shows the averaged values of violent crimes during 2013-2017 in Tokyo. The plot indicates that the distribution of violent crimes has spatial heterogeneity and there are several hotspots. \cite{hamura2021robust} regarded the hotspots as outliers and proposed a robust method for violent crimes in 2017. On the other hand, \cite{yano2021minimax} focused on the pickpocket (not violent crime) from 2012 to the first half of 2018 at 978 towns in eight wards, and considered a Bayesian prediction problem based on the Poisson distribution. 
Our goal in this paper is to estimate the spatial high-risk trends of violent crime with uncertainty and to detect the potential risk throughout multiple years simultaneously. It is important to adaptively estimate trends without smoothing for potentially high-risk areas. 

Spatial data with longitude and latitude information are considered point-level data. Statistical methods for such data have been developed. As a nonparametric Bayesian approach, \cite{taddy2010autoregressive} considered the autoregressive mixture model and also provided an application of crime data analysis. On the other hand, data observed per region is known as areal data. Tokyo crime data considered in this paper is areal data, and it is constructed by the total number of crimes per region for a year. In other words, it does not make much sense to consider it as point-level data. For crime data as areal data, \cite{balocchi2019spatial} proposed a Bayesian linear regression model over time within a spatial correlation like conditional autoregressive formulation, and they applied their method to an analysis of violent crimes in Philadelphia. For the same data, \cite{balocchi2023crime} also proposed the CAR-within-clusters model which assumes linear formulation and conditionally autoregression (CAR) model for each cluster, which deals with spatial discontinuity by introducing cluster and gives spatial continuity within a cluster. They recommend using the {\it crime density} defined as the number of crimes divided by the land area to deal with the difference in land size. The approach treats crime data as a continuous value instead of count data. Following the study, we adopt the crime density in the Tokyo crime data, that is, we assume the continuous distribution on the distribution generating data in our modeling, not count data such as Poisson distribution.

In this paper, we develop a quantile trend estimation for spatial data. The smoothing method has been studied in the context of function estimation to investigate the characteristics of the time series data. The $\ell_1$ trend filtering \citep{kim2009ell_1,tibshirani2014adaptive} is a nonparametric method to estimate underlying trends, which archives locally adaptive smoothing compared with spline methods, and fast and efficient optimization algorithms were also proposed \citep[e.g.][]{ramdas2016fast}. For these reasons, extensions of the original trend filtering have been considered, such as the trend filtering on graphs \citep{wang2015trend} and for functional data \citep{wakayama2023trend}. The Bayesian formulation of trend filtering based on Gaussian likelihood and shrinkage priors has been considered \citep[e.g.][]{roualdes2015bayesian,faulkner2018locally, heng2023bayesian}, and the extension to dynamic shrinkage process was also proposed by \cite{kowal2019dynamic}. 
While these methods focus on mean trends, to estimate quantile trends instead of mean, \cite{brantley2020baseline} proposed quantile trend filtering, which was compared with the spline method and provides reasonable estimates of the baseline even under the presence of outliers. As the Bayesian methods, \cite{onizuka2022fast} and \cite{barata2022fast} proposed the Bayesian quantile trend filtering and the extended dynamic quantile linear model for time series data, respectively. By accounting for covariates, \cite{reich2011bayesian} proposed a Bayesian spatial quantile regression by introducing spatially varying basis-function coefficients. 
\cite{castillo2023spatial} also considered spatial quantile autoregression for space-time dependence data, which is based on the Gaussian process model to capture spatial dependence over the grid cells.

There are some difficulties with these methods. The main difficulty in applying frequentist trend filtering is that uncertainty quantification is not straightforward. Moreover, the frequentist formulation includes tuning parameters that influence smoothness in the penalty term, but the data-dependent selection of the tuning parameter is not obvious,  especially under quantile smoothing. While Bayesian methods are capable of mitigating these issues, the existing approach only focuses on time series data; thereby it cannot handle the smoothing of data on general graphs such as spatial data. 
Moreover, most of the studies focused on estimating mean trend under a homogeneous variance structure, and these methods may not work well in data with heterogeneous variance. Nevertheless, quantile smoothing for spatial data has not been studied even from a frequentist perspective.

\begin{figure}[thbp]
\begin{center}
\includegraphics[width=\linewidth]{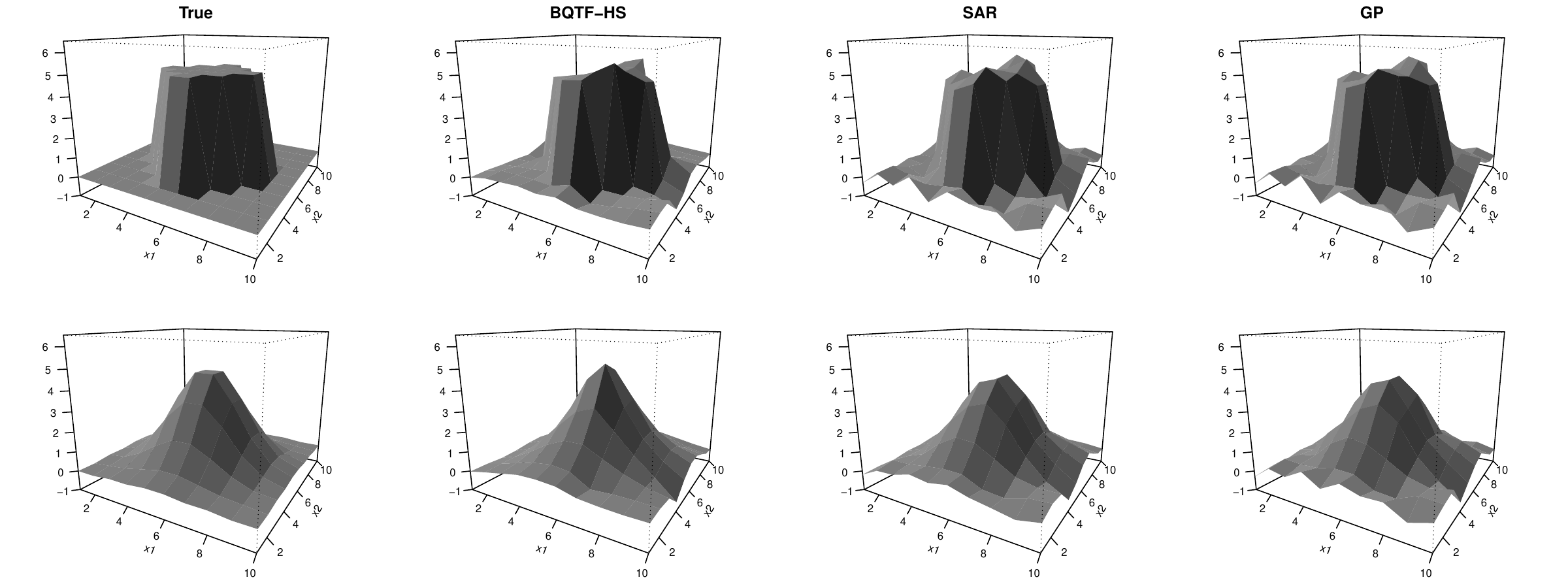}
\caption{The examples of three methods for the 0.5-th quantile level. The left panels are two true signals. The estimates of the proposed methods under horseshoe and $k=1$, SAR models, and GP models for two signals from the second from left to right.}
\label{Oneshot-example-4methods}
\end{center}
\end{figure}

To overcome the issues, we extend the Bayesian quantile smoothing for time series data to Bayesian quantile trend filtering on general graphs including spatial neighboring structures, and also allow for multiple spatial data in which the number of samples for each location may be different. 
To this end, we employ the asymmetric Laplace distribution as a working likelihood \citep{yu2001bayesian}, where the theoretical justification of using the likelihood is discussed in \cite{sriram2013posterior} and \cite{sriram2015sandwich}.
The novelty of the proposed approach is the construction of the prior distribution on the graph difference. 
In particular, we consider the horseshoe prior \citep{carvalho2010horseshoe} as locally adaptive shrinkage priors for the graph differences. 
We introduce a novel hierarchical formulation for the prior, known as ``shadow priors" that enhances the efficiency of posterior computation. 
Specifically, combining the data augmentation strategy by \cite{kozumi2011gibbs}, we develop a simple Gibbs sampling algorithm to generate posterior samples. We demonstrate the usefulness and wide applicability of proposed methods through extensive simulation studies and application to Tokyo crime data. 
We here present the advantage of the proposed trend filtering method compared with the existing Bayesian spatial methods: the simultaneous autoregressive (SAR) model and the Gaussian process (GP) model. 
In Figure~\ref{Oneshot-example-4methods}, we show two examples of true quantile trends (adopted in simulation studies in Section~\ref{sec:3}), and their estimated results obtained by the proposed method (BQTF-HS) as well as SAR and GP models.
It is observed that BQTF-HS tends to provide better estimation results than both SAR and GP models, successfully taking account of local changes and the smoothness of the true trend. 
Note that similar advantages of trend filtering were confirmed in the context of smoothing mean parameters \citep{tibshirani2014adaptive,wang2015trend}.

The paper is organized as follows: In Section \ref{sec:2}, we propose a new Bayesian trend filtering method to estimate quantiles and construct an efficient posterior sampling algorithm based on Gibbs sampling. In Section \ref{sec:3}, we illustrate some simulation studies to compare the performance of proposed methods. In Section \ref{sec:4}, we apply the proposed methods to violent crime data in Tokyo.
Additional numerical results are provided in the Supplementary Material. {\tt R} code implementing the proposed methods is available in the GitHub repository (URL: \url{https://github.com/Takahiro-Onizuka/BSQS}).

\section{Bayesian quantile trend filtering on graphs}
\label{sec:2}

\subsection{Background}
\label{subsec:2.1}

Let $y_i = \theta_i +\varepsilon_i\quad (i=1,\dots,n)$ be a sequence model, where $y_i$ is an observation, $\theta_i$ is a true function and $\varepsilon_i$ is a noise. Let $\hat{\theta}$ be the minimizer of the following penalized problem:
\begin{align}\label{loss-pen}
\hat{\theta}=\arg \min_{\theta\in\mathbb{R}}  \ell(y-\theta)+\lambda\|D^{(k+1)}\theta\|_1 ,
\end{align}
where $\ell(\cdot)$ is a loss function, $y=(y_1,\dots,y_n)^{\top}$, $\theta=(\theta_1,\dots,\theta_n)^{\top}$, $D^{(k+1)}$ is a $(n-k-1)\times n$ difference operator matrix of order $k+1$, and $\lambda>0$ is a tuning constant. For $\ell(y-\theta)=\|y-\theta\|_2^2$ in \eqref{loss-pen}, the optimization problem corresponds to $\ell_1$ trend filtering \citep{kim2009ell_1}. We note that the $\ell_1$ trend filtering is considered as a special case of the generalized lasso proposed by \cite{tibshirani2011solution}. From a computational perspective, \cite{ramdas2016fast} proposed a fast and efficient optimization algorithm to obtain the trend filtering estimate. Depending on the different order $k$, we can express various smoothing such as piecewise constant, linear, quadratic, and so on \citep{tibshirani2014adaptive}. For spatial data, the trend filtering on graphs was also proposed by \cite{wang2015trend} based on the graph difference operator instead of the standard difference operator in \eqref{loss-pen}. 

Recently, different loss functions are also considered. For example, \cite{brantley2020baseline} considered the check loss function $\ell(y-\theta)=\rho_p(y-\theta)$, and proposed quantile trend filtering to estimate the trend in the baseline, not the mean. To solve the problem, \cite{brantley2020baseline} proposed a parallelizable alternating direction method of multipliers (ADMM) algorithm. Furthermore, they also provided a modified criterion based on the extended Bayesian information criterion to select the tuning parameter. 

We next introduce Bayesian trend filtering. In general, the Bayesian formulation for trend filtering is based on the model:
\begin{align}
    y_i=\theta_i+\varepsilon_i,\quad \varepsilon_i\sim f(\cdot), \quad D^{(k+1)}\theta\sim \pi(\cdot),\quad (i=1,\ldots,n),
    \label{BTFseq}
\end{align}
where $f$ and $\pi$ correspond to the likelihood and prior density functions, respectively. A simple Bayesian counterpart that corresponds to penalized square loss is a combination of the Gaussian likelihood on $f$ and Laplace prior distribution on $\pi$ \citep[e.g.][]{roualdes2015bayesian}. The resulting posterior mode is the same as that of the solution of the problem \eqref{loss-pen}. However, it is well-known that the shrinkage based on Laplace prior often causes over-shrinkage due to the tail of Laplace distribution. Recently, \cite{faulkner2018locally} proposed a more flexible Bayesian trend filtering via global-local shrinkage priors such as horseshoe prior \citep{carvalho2010horseshoe}. Assuming asymmetric Laplace likelihood in \eqref{BTFseq}, \cite{onizuka2022fast} proposed a Bayesian quantile trend filtering. They also provided a calibrated variational Bayes algorithm to reduce the misspecification bias of asymmetric Laplace likelihood. \cite{barata2022fast} consider the model \eqref{BTFseq}, but they employed a more flexible probability distribution called extended asymmetric Laplace distribution.

\subsection{Shrinkage priors on graph differences}
\label{subsec:2.2}

Following \cite{onizuka2022fast}, we will consider the following model:
\begin{align}
&y_{ij}=\theta(x_i)+\varepsilon_{ij},\quad \varepsilon_{ij}\sim \mathrm{AL}(p,\sigma^2), \ \ \ i=1,\ldots,n, \quad j=1,\dots, N_i, \label{ALseq}
\end{align}
where $y_{ij}$ is a $j$th observation in the location $x_i$, $\theta(x_i)=\theta_i$ is a common quantile to $y_{i1},\dots,y_{iN_i}$ in the location $x_i$, $N_i$ is the number of data per each location $x_i$, $\sigma^2$ is unknown parameters, and $p$ is a fixed quantile level.
Here ${\rm AL}(p,\sigma^2)$ denotes the asymmetric Laplace distribution:
\begin{align*}
f_p(x)=\frac{p(1-p)}{\sigma^2}\exp\left\{-\rho_p\left(\frac{x}{\sigma^2}\right)\right\},
\end{align*}
where $p$ is a fixed constant which characterizes the quantile level, $\sigma^2$ (not $\sigma$) is a scale parameter and $\rho_p(\cdot)$ is a check loss function given by 
\begin{align*}
\rho_p(r)=\sum_{i=1}^nr_i\{ p-1(r_i<0) \}, \quad 0<p<1.
\end{align*}
Note that the model \eqref{ALseq} handles a situation with multiple observations per grid point, and $(y_{i1},\ldots,y_{iN_i})$ are marginally correlated due to the common $\theta(x_i)$.

Suppose that spatial location $x=(x_1,\dots,x_n)$ has a graph structure, and then $\theta_1,\ldots,\theta_n$ are on general graphs (including the standard trend filtering as a linear chain graph). The assumption is commonly used because the areal data has an adjacency relation and the simultaneous/conditional autoregressive models are also based on graph structure. 
Following \cite{wang2015trend}, let $G=(V,E)$ be an undirected graph with vertex set $V=\{1,\dots,n\}$ and edge set $E$. 
We assume that $|V|=n$ and $|E|=m$. For $k=0$, if $e_{\ell}=(i,j)\in V$, then $D^{(1)}$ has $\ell$-th row
\begin{align}
     D^{(1)}_{\ell}=(0,\dots,0,\underbrace{1}_{i},0\dots,0,\underbrace{-1}_{j},0,\dots,0),
     \label{1-difference operator}
\end{align}
where $1\leq \ell\leq m$. For a graph $G$, the graph difference operator of order $k+1$ is denoted by $D^{(k+1)}$. When $k\geq 1$, graph difference operator $D^{(k+1)}$ is defined by
\begin{align}
D^{(k+1)}=
\begin{cases}
(D^{(1)})^{\top}D^{(k)} & \mathrm{ for\ odd\ } k,\\
D^{(1)}D^{(k)} & \mathrm{ for\ even\ } k.
\end{cases}\label{k-difference operator}
\end{align}
Here, we have $D^{(k+1)}\in \mathbb{R}^{n\times n}$ for odd $k$ and $D^{(k+1)}\in \mathbb{R}^{m\times n}$ for even $k$. We note that the first-order graph difference operator $D^{(1)}$ is a natural generalization of the usual first-order difference operator used in \cite{kim2009ell_1}, and if we consider the linear chain graph corresponding to time series data, then they coincide. 
The $k$ controls the smoothness of the estimated trend. For example, $k=0$ represents the assumption that the trend to estimate is piecewise constant like the upper left in Figure~\ref{Oneshot-example-4methods}. $k\ge 1$ corresponds to the piecewise polynomial trend with degree $k$ as an estimate for the unknown spatial trend. In other words, the estimate of $\theta_i$ has a relationship with its neighboring values like a polynomial function, which is similar to a local linear/polynomial regression. Empirically, we recommend $k=1$ to capture changes and avoid over-fitting.

Let $D$ be a $m\times n$ full-rank matrix representing a general difference operator on a graph, and we consider flexible shrinkage priors on $D\theta$. 
When $m$ is smaller than $n$ as in a linear chain graph, $D$ can be transformed to $n\times n$ non-singular matrix \citep[see also][]{onizuka2022fast}. We here assume that $m\geq n$ since the number of edges is typically larger than that of nodes. 
We consider the prior $D\theta\mid \tau^2,\sigma^2,w\sim N_n(0,\tau^2\sigma^2 W)$ with a diagonal covariance matrix $W={\rm diag}(w_1^2,\ldots,w_m^2)$, where $w=(w_1,\ldots,w_m)$ represents local shrinkage parameters for each element in $D\theta$ and $\tau^2$ is a global shrinkage parameter. 
When $m=n$, the prior can be rewritten as 
\[
\theta\mid\tau^2,\sigma^2,w \sim N_n(0,\sigma^2\tau^2(D^{\top}W^{-1} D)^{-1}).
\]
Our idea is to use the above prior form even under $m>n$, noting that the covariance matrix $(D^{\top}W^{-1} D)^{-1}$ is still non-singular under $m>n$. 
The density function of the conditional prior of $\theta$ is given by 
\begin{equation}\label{prior-theta}
\pi(\theta\mid \tau^2,\sigma^2,w)=(2\pi\sigma^2\tau^2)^{-n/2}|D^{\top}W^{-1} D|^{1/2}\exp\left(-\frac1{2\sigma^2\tau^2}\theta^{\top}D^{\top}W^{-1} D\theta\right).
\end{equation}
Now, we consider the prior for $w$. 
The standard approach is the use of an independent prior $\pi(w)=\prod_{i=1}^m \pi(w_i)$, and some familiar distribution is used for $\pi(w_i)$, for example, exponential prior or inverse gamma prior. 
However, the full conditional distribution of $w$ is not a familiar form due to the term $|D^{\top}W^{-1} D|^{1/2}$ in the density (\ref{prior-theta}). Therefore, it is not easy to construct an efficient Gibbs sampler. 
Alternatively, we consider the following joint prior: 
\begin{equation}\label{prior-w}
\pi(w)\propto |D^{\top}W^{-1} D|^{-1/2}|W|^{-1/2}\prod_{i=1}^m\pi(w_i),
\end{equation}
where $\pi(w_i)$ is a proper univariate distribution. For a square matrix $D$ such that $k$ is odd, the joint prior equals the product of the standard prior $\pi(w_i)$. As shown in Subsection \ref{subsec:2.3}, the resulting full conditional distributions of $w$ are familiar forms under well-known priors for local shrinkage parameters. As a result, we can construct a Gibbs sampler for the proposed method. 
Such priors given in (\ref{prior-w}) are known as ``shadow priors", and are used to improve the mixing of Markov chain Monte Carlo (MCMC) algorithm \citep[e.g.][]{liechty2009shadow} or to construct tractable full conditional distributions \citep[e.g.][]{liu2014bayesian, xu2015bayesian}.
Note that these works demonstrate that the use of shadow prior has little effect on posterior inference.

As an univariate distribution $\pi(w_i)$ in (\ref{prior-w}), we consider two types of distributions, $w_i\sim {\rm Exp}(1/2)$ and $w_i\sim C^{+}(0,1)$. These priors are motivated by the Bayesian lasso prior \citep{park2008bayesian} and horseshoe prior \citep{carvalho2010horseshoe}, respectively. 
Regarding the other parameters, we assign $\sigma^2\sim {\rm IG}(a_{\sigma}, b_{\sigma})$ and $\tau \sim C^{+}(0, C_{\tau})$, where $a_{\sigma}$, $b_{\sigma}$ and $C_{\tau}$ are fixed hyper-parameters.

The proposed prior for $\theta$ belongs to a class of general priors, described as
\begin{align}\label{SARprior}
\theta\mid \sigma^2, \tau^2, \rho \sim N(0, \sigma^2\tau^2 Q(\rho)).
\end{align}
Note that the simultaneous autoregressive (SAR) and Gaussian process (GP) prior are popular approaches for spatial smoothing and the priors can also be expressed as (\ref{SARprior}) with different matrix $Q(\rho)$ from that of the proposed prior. 
The two priors will be compared through simulation studies and more detailed explanations are provided in Section \ref{sec:3}.

Note that the three conditional priors of $\theta$ include $\sigma^2$ in the scale although $\sigma^2$ is the scale parameter of the likelihood (see equations \eqref{prior-theta} and \eqref{SARprior}). The formulation has been often used for the conditional normal prior \citep[e.g.][]{polson2012half} and induces the advantage that the scale of the prior is automatically adjusted when units of observations are changed.

\subsection{Markov chain Monte Carlo algorithm}
\label{subsec:2.3}

To develop an efficient posterior computation algorithm via Gibbs sampling, we employ the stochastic representation of the asymmetric Laplace distribution \citep{kozumi2011gibbs}.
For $\varepsilon_{ij} \sim \mathrm{AL}(p,\sigma^2)$, we have the following argumentation
\[
\varepsilon_{ij}=\psi z_{ij}+\sqrt{\sigma^2 z_{ij} t^2}u_{ij},\quad \psi=\frac{1-2p}{p(1-p)},\quad t^2=\frac{2}{p(1-p)},
\]
where $u_{ij} \sim N(0,1)$ and $z_{ij}\mid \sigma^2 \sim \mathrm{Exp}(1/\sigma^2)$ for $i=1,\ldots,n$. 
From the above expression, the conditional likelihood function of $y_{ij}$ is given by
\[
p(y_{ij}\mid \theta_i, z_{ij}, \sigma^2)=(2\pi t^2\sigma^2)^{-1/2}z_{ij}^{-1/2}\exp\left\{ -\frac{(y_{ij}-\theta_i-\psi z_{ij})^2}{2t^2\sigma^2z_{ij}}\right\}.
\]
Then, under the conditionally Gaussian prior of $\theta$ in (\ref{prior-theta}), the full conditional distributions of $z_i$ and $\theta$ are given by 
\begin{align*}
&\theta\mid y,z,\sigma^2,\gamma^2
\sim N_n\left(A^{-1}B, \sigma^2 A^{-1}\right),\\
&z_{ij}\mid y_{ij},\theta_i,\sigma^2\sim \mathrm{GIG}\left(\frac{1}{2},\frac{(y_{ij}-\theta_i)^2}{t^2\sigma^2},\frac{\psi^2}{t^2\sigma^2}+\frac{2}{\sigma^2}\right), \ \ i=1,\ldots,n, \ j=1,\dots,N_i,
\end{align*}
where 
\begin{align*}
    A&=\frac{1}{\tau^2}D^{\top}W^{-1}D+\frac{1}{t^2}\diag\left(\sum_{j=1}^{N_1}z_{1j}^{-1},\ldots,\sum_{j=1}^{N_n}z_{nj}^{-1}\right), \\
    B&=\left(\sum_{j=1}^{N_1}\frac{y_{1j}-\psi z_{1j}}{t^2z_{1j}},\dots,\sum_{j=1}^{N_n}\frac{y_{nj}-\psi z_{nj}}{t^2z_{nj}}\right)^{\top}
\end{align*}
and $\mathrm{GIG}(a,b,c)$ denotes the generalized inverse Gaussian distribution.
The full conditional distributions of the scale parameter of observations, $\sigma^2$, and global shrinkage parameter $\tau^2$ are given by 
\begin{align*}
&\sigma^2\mid y,\theta, z, w,\tau^2\sim \mathrm{IG}\left(\frac{n+3N}{2}+a_{\sigma},
\beta_{\sigma^2}
\right),\\
&\tau^2\mid \theta, w, \sigma^2,\xi\sim \mathrm{IG}\left(\frac{n+1}{2}, \frac1{2\sigma^2}\theta^\top D^\top W^{-1}D\theta+\frac{1}{\xi}\right), \ \ \ \ \
\xi\mid \tau^2\sim \mathrm{IG}\left(\frac{1}{2},\frac{1}{\tau^2}+1\right),\\
&\beta_{\sigma^2}=\sum_{i=1}^n\sum_{j=1}^{N_i}\frac{(y_{ij}-\theta_i-\psi z_{ij})^2}{2t^2z_{ij}}+\frac{\theta^\top D^\top W^{-1}D\theta}{2\tau^2}+\sum_{i=1}^n\sum_{j=1}^{N_i} z_{ij}+b_{\sigma},
\end{align*}
where $N$ is the number of total data and $\xi$ is an augmented parameter for $\tau^2$.
The full conditional distributions of the other parameters depend on the specific choice of the distributional form of $\pi(w_i)$, which are summarized as follows. 

\begin{itemize}

\item[-]{\bf (Laplace-type prior)}  \ \ 
The full conditional distributions of $\theta$, $z_i$, and $\sigma^2$ have already been mentioned. For the Laplace-type prior, we give $\tau^2=1$ and $w_i\mid \gamma^2 \sim \mathrm{Exp}(\gamma^2/2)$. In this condition, we can model that $(D\theta)_i \sim \mathrm{Lap}(\gamma)$. Because our condition is $\gamma \sim C^+(0,1)$, by using the representation that if $\mathrm{IG}(\gamma^2\mid 1/2,1/\nu)$ and $\mathrm{IG}(\nu\mid 1/2, 1/a^2)$, then $\gamma\sim C^+(0,a)$, the full conditional distributions of $w_i$, $\gamma^2$ and $\nu$ are given by 
\begin{align*}
&w_i^2\mid \theta,\sigma^2,\gamma^2,\nu \sim \mathrm{GIG}\left(\frac{1}{2},\frac{\eta_i^2}{\sigma^2},\gamma^2\right),\\
&\gamma^2\mid w, \nu\sim \mathrm{GIG}\left(m-\frac{1}{2},\frac{2}{\nu},\sum_{i=1}^m w_i^2\right),\quad \nu\mid \gamma^2 \sim \mathrm{IG}\left(\frac{1}{2},\frac{1}{\gamma^2}+1\right),
\end{align*}
where $\mathrm{GIG}(a,b,c)$ is the generalized inverse Gaussian distribution and $\eta_i=(D\theta)_i$.

\item[-]{\bf (Horseshoe-type prior)} \ \ 
The full conditional distributions of $\theta$, $z_i$, $\sigma^2$ and $\tau^2$ have already been mentioned. For the Horseshoe-type prior, $w_i \sim C^+(0,1)$. By using the representation that  $w_i^2\mid \nu_i \sim \mathrm{IG}(1/2,1/\nu_i)$ and $\nu_i\sim \mathrm( 1/2, 1)$, the full conditional distributions of $w_i$ and $\nu_i$ are given by 
\begin{align*}
&w_i^2\mid \theta,\sigma^2,\gamma^2,\nu \sim \mathrm{IG}\left(1,\frac{1}{\nu_i}+\frac{\eta_i^2}{2\sigma^2\tau^2}\right),\quad \nu_i\mid w_i \sim \mathrm{IG}\left(\frac{1}{2},\frac{1}{w_i^2}+1\right),
\end{align*}
where $\mathrm{IG}(a,b)$ is the inverse Gamma distribution and $\eta_i=(D\theta)_i$.

\end{itemize}


\section{Simulation studies}
\label{sec:3}

We illustrate the performance of the proposed method through simulation studies. 

\subsection{Simulation setting}
\label{subsec:3.1}

\begin{figure}[thbp]
\begin{center}
\includegraphics[width=0.8\linewidth]{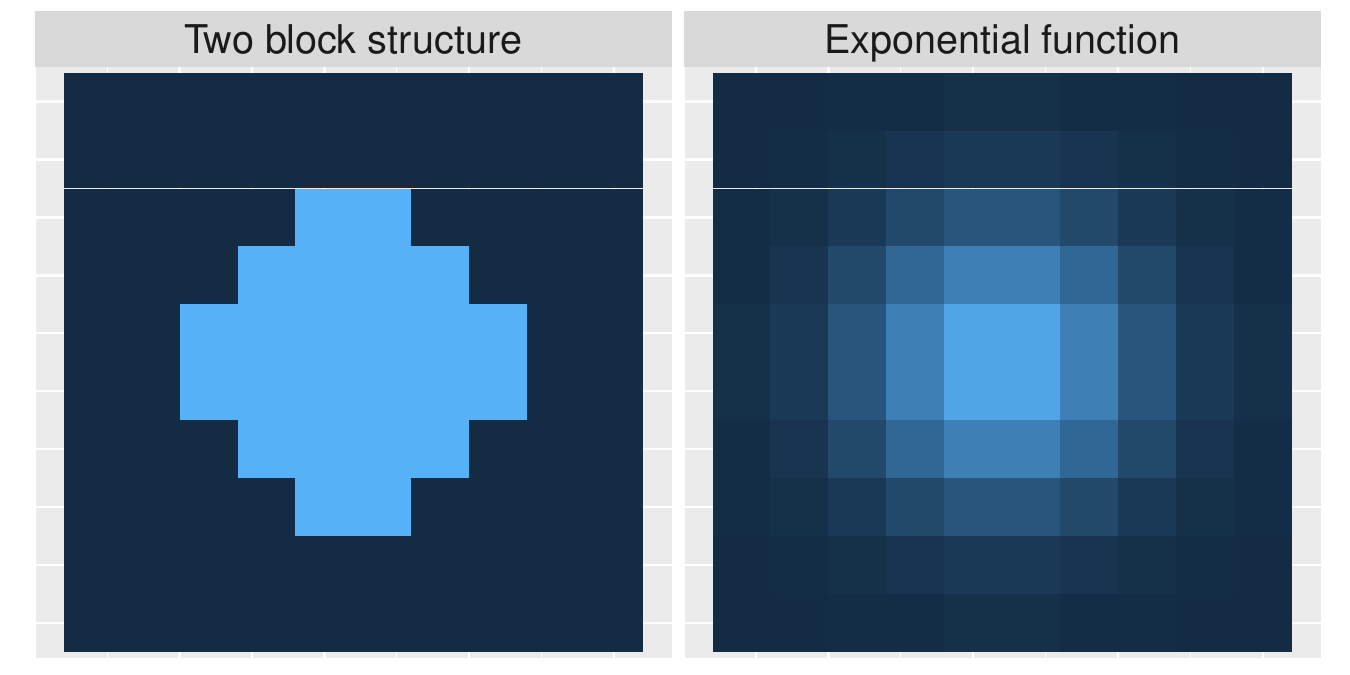}
\caption{Two types of true function $f(x)$.}
\label{2D_true_function}
\end{center}
\end{figure}

We show simulation studies for data on 2-D lattice graphs. 
We formulate the data-generating process as follows. Let $G=(V,E)$ be a 2-D lattice graph. We set $V=\{1,\dots,100\}$ and $|E|=180$ for the graph $G$. The edges are defined by whether the lattice is adjacent or not. A more general graph structure is also considered in the Supplementary Material. Noisy data were generated from the model $y_{ij}=f(x_i)+\varepsilon(x_i)$ ($i=1,\dots,100$, $j=1,\dots,5$), where $x_i=(x_{i1}, x_{i2})$ is a two-dimensional coordinate, and $f(x)$ and $\epsilon(x)$ are true and noise functions, respectively. Based on the model \ref{ALseq}, we generated five data for each location $i$. The following two true functions were considered:
\begin{itemize}
\item Two block structure
$$
f(x_i)=5 \ \  (\text{center}), \ \ \ \  \text{and} \ \ \ \ 
f(x_i)=0 \ \  (\text{other}),
$$
\item Exponential function
\begin{align*}
f(x_i)&=5\exp\left(-\frac{1}{2}(x_i-\mu)^{\top}\Sigma^{-1}(x_i-\mu)\right),\quad \mu=(5.5, 5.5),\quad \Sigma=3I_2,
\end{align*}
\end{itemize}
where $x_i=(x_{i1},x_{i2})$, $x_{i1},x_{i2}=1,2,\dots,10$ and $I_n$ is $n\times n$ identity matrix. These functions are shown in Figure \ref{2D_true_function}. 
As noise functions $\epsilon(x)$, we considered the following three structures:
\begin{itemize}
\item[(I)] Homogeneous: $\epsilon(x_i) \sim N(0,1)$.
\item[(II)] Block heterogeneous:
\begin{align*}
\epsilon(x_i) \sim 
\begin{cases}
N(0,0.5^2) & (1\le x_{i1} \le 5,\ 1\le x_{i2} \le 5)\\
N(0,2^2) & (6\le x_{i1} \le 10,\ 6\le x_{i2} \le 10)\\
N(0,1) & (\text{otherwise})
\end{cases} .
\end{align*}
\item[(III)] Smooth heterogeneous:
\begin{align*}
\epsilon(x_i) \sim 
\begin{cases}
N(0,0.5^2) & (1\le x_{i1} \le 4,\ 1\le x_{i2} \le 4)\\
N(0,1) & (x_{i1}=5,6,\ 1\le x_{i2} \le 6\ \text{or}\ 1\le x_{i1} \le 6,\ x_{i2}=5,6)\\
N(0,1.5^2) & (x_{i1}=7,8,\ 1\le x_{i2} \le 8\ \text{or}\ 1\le x_{i1} \le 8,\ x_{i2}=7,8)\\
N(0,2^2) & (\text{otherwise})
\end{cases} .
\end{align*}
\end{itemize}
The two-block structure is a reasonable function to verify the ability to capture the jump without smoothing. In the exponential function, we examine the ability to estimate a continuous curve with noisy data. 
The noise (I) represents spatial homogeneity, while the noise may not be realistic in practical situations. In noise (II) and (III), the aim is to verify how well the proposed method can handle spatial heterogeneity. In particular, noise (III) has a stronger degree of spatial heterogeneity than noise (II). The visualizations of these noise distributions are given in the Supplementary Material. 
Combining two true structures and three noise functions, we consider six scenarios. Scenarios (i), (ii), and (iii) are based on two block structure and noise functions (I), (II), and (III), respectively. Scenarios (iv), (v), and (vi) are based on exponential structure and noise functions (I), (II), and (III), respectively.
Since the noise function is homogeneity, scenarios (i) and (iv) are easier, and scenario (iv) would especially be the easiest because of its smoothness and homogeneity. Scenario (iii) constructed by a two-block structure and smooth heterogeneous noise would be the most difficult for two reasons: hard to capture the jump points and heavy heterogeneity.

We used the two proposed methods (denoted by BQTF-HS and BQTF-Lap), where HS and Lap are the horseshoe and Laplace priors, respectively. Although there is no previous research about spatial quantile smoothing, to evaluate the performance of the proposed method, we compare the BQTF methods with the following three methods:
\begin{itemize}
\item 
SAR: Bayesian simultaneous autoregressive (SAR) quantile model, which is based on graph structure as well as BQTF. The SAR prior takes the form of \eqref{SARprior} and is based on a graph structure with a contingency matrix $\Omega$. The matrix $Q(\rho)$ is given by
$Q(\rho)^{-1}=(I_n-\rho \Omega)^\top (I_n-\rho \Omega)$, and the parameter $\rho$ controls the effect of the spatial correlation.
The MCMC algorithm is summarized in the Supplementary Material. Note that the parameter $\rho$ was sampled by the random walk Metropolis-Hastings (MH) algorithm.

\item 
GP: Bayesian quantile smoothing under Gaussian process prior, which the prior takes the form of \eqref{SARprior} and is based on the location of a data observed point. The matrix $Q(\rho)$ is given by $Q_{ij}(\rho):=(Q(\rho))_{i,j}=\exp\left(-\|x_i-x_j\|/(2\rho)\right)$, where $\rho$ also controls the effect of the spatial dependence between the location $x_i$ and $x_j$.
The MCMC algorithm is summarized in the Supplementary Material. Note that the parameter $\rho$ was sampled by the random walk MH algorithm. 

\item 
qgam: Additive quantile regression which is the frequentist method proposed by \cite{fasiolo2021fast}.
The method can be implemented by using their \texttt{qgam} R package. 
Let $x_i=(x_{i1}, x_{i2})$ be a two-dimensional coordinate and $y_i$ be a observed data. Then the corresponding estimate $\hat{\theta}_i$ of $\theta_i$ is obtained by the sum of functions $\hat{\theta}_i=\hat{q}_1(x_{i1})+\hat{q}_2(x_{i2})$, where $\hat{q}_1(\cdot)$ and $\hat{q}_2(\cdot)$ are nonparameteric estimates of quantile functions.
\end{itemize}
Note that the detailed posterior computation algorithms of SAR and GP methods are presented in the Supplementary Material. 
Since the SAR and GP models are based on the Gaussian type prior, we could not expect locally adaptive smoothing. If the data $y_i$ is generated from a simple true function $f(x_i)=f_1(x_{i1})+f_2(x_{i2})$, then the qgam method would give a pretty good smoothing, but it is unrealistic in a practical situation and the simulation setting is more complicated. The other methods such as spatial regression models are also compared and the results are summarized in the Supplementary Material. 
For the Bayesian methods, we generated 7,500 posterior samples, and then only every 10th scan was saved, and the order of trend filtering was set as $k=1$ (i.e. area-wise linear trend). We estimate five quantile levels: 0.1, 0.3, 0.5, 0.7, and 0.9. To evaluate the performance of estimates, we adopt the mean squared error (MSE), the mean absolute deviation (MAD), the mean credible interval width (MCIW), and the coverage probability (CP) which are defined by
\begin{align*}
    &\mathrm{MSE}=\frac{1}{n}\sum_{i=1}^n (\theta_i^*-\hat{\theta}_i)^2,\quad \mathrm{MAD}=\frac{1}{n}\sum_{i=1}^n \vert \theta_i^*-\hat{\theta}_i\vert ,\\ &\mathrm{MCIW}=\frac{1}{n}\sum_{i=1}^n \hat{\theta}_{97.5,i}-\hat{\theta}_{2.5,i},\quad
    \mathrm{CP}=\frac{1}{n}\sum_{i=1}^n I(\hat{\theta}_{2.5,i}\leq \theta_i^* \leq \hat{\theta}_{97.5,i}),
\end{align*}
respectively, where $\hat{\theta}_{100(1-\alpha),i}$ represent the $100(1-\alpha)$\% posterior quantiles of $\theta_i$ and $\theta_i^*$ are true quantiles at location $x_i$. These values were averaged over 100 replications of simulating datasets.

\subsection{Simulation result}
\label{subsec:3.2}

Simulation results are shown in Tables \ref{table_1} and \ref{table_2}. Note that MCIW and CP are reported only for Bayesian methods. From Tables \ref{table_1} and \ref{table_2}, the proposed BQTF method under horseshoe prior tends to provide a reasonable point estimate not only in the case of homogeneous but also for heterogeneous variances. 
When the true structure is the exponential function (such as scenarios (iv), (v), and (vi)), the proposed two methods provide comparable point estimates, and the additive quantile regression has smaller MSE and MAD than that for (i), (ii) and (iii) scenarios. However, it is observed that the additive quantile regression does not work well for any scenario compared with the proposed methods. 
The MAD of the proposed BQTF-HS is smaller than that of the SAR and GP models for all cases, while the SAR model is sometimes the best for exponential function in terms of MSE because of the smooth trend structure. In comparison between the SAR and GP models, the SAR model is better than the GP model in terms of MSE and MAD.
For uncertainty quantification, while BQTF methods have reasonable coverage probabilities for the 50\% quantile trend, the coverage probabilities of 95\% credible intervals for extremal quantiles such as 0.1 and 0.9 seem to be far away from the nominal coverage rate of 0.95. Note that the mean credible interval width (MCIW) is the order of HS, Lap, SAR, and GP.

\begin{table}[thbp]
\begin{center}
\caption{Average values of MSE, MAD, MCIW, and CP based on $100$ replications for scenarios (i), (ii), and (iii) (two-block structure). The minimum values of MSE and MAD are represented in bold. }
\resizebox{1.0\textwidth}{!}{ 
\begin{tabular}{|c|ccccc|ccccc|}
\hline
&\multicolumn{10}{c|}{Scenario (i)}\\
\hline
&\multicolumn{5}{c|}{MSE}&\multicolumn{5}{c|}{MAD}\\
\hline
 &0.1& 0.3 & 0.5 & 0.7 &0.9&0.1& 0.3 & 0.5 & 0.7 &0.9\\
\hline
  HS & {\bf 0.264} & {\bf 0.158} & {\bf 0.138} & {\bf 0.152} & {\bf 0.248} & {\bf 0.377} & {\bf 0.283} & {\bf 0.266} & {\bf 0.280} & {\bf 0.371} \\ 
  Lap & 0.337 & 0.212 & 0.193 & 0.211 & 0.339 & 0.460 & 0.359 & 0.344 & 0.360 & 0.463 \\ 
  SAR & 0.345 & 0.224 & 0.206 & 0.224 & 0.345 & 0.469 & 0.377 & 0.362 & 0.378 & 0.470 \\ 
  GP & 0.347 & 0.217 & 0.199 & 0.217 & 0.354 & 0.470 & 0.372 & 0.356 & 0.373 & 0.476 \\ 
  qgam & 3.985 & 2.461 & 1.877 & 2.269 & 3.062 & 1.256 & 1.266 & 1.195 & 1.266 & 1.425 \\ 
  \hline
  &\multicolumn{5}{c|}{MCIW}&\multicolumn{5}{c|}{CP}\\
\hline
 & 0.1 & 0.3 & 0.5 & 0.7 & 0.9 & 0.1 & 0.3 & 0.5 & 0.7 & 0.9 \\ 
  \hline
  HS & 1.322 & 1.287 & 1.262 & 1.261 & 1.290 & 0.822 & 0.920 & 0.934 & 0.921 & 0.827 \\ 
  Lap & 1.400 & 1.554 & 1.552 & 1.541 & 1.395 & 0.799 & 0.910 & 0.921 & 0.906 & 0.801 \\ 
  SAR & 1.539 & 1.706 & 1.706 & 1.693 & 1.534 & 0.826 & 0.923 & 0.935 & 0.920 & 0.829 \\ 
  GP & 1.525 & 1.712 & 1.717 & 1.703 & 1.510 & 0.826 & 0.928 & 0.942 & 0.927 & 0.826 \\ 
\hline
\hline
&\multicolumn{10}{c|}{Scenario (ii)}\\
\hline
&\multicolumn{5}{c|}{MSE}&\multicolumn{5}{c|}{MAD}\\
\hline
 &0.1& 0.3 & 0.5 & 0.7 &0.9&0.1& 0.3 & 0.5 & 0.7 &0.9\\
\hline
  HS & {\bf 0.464} & {\bf 0.272} & {\bf 0.244} & {\bf 0.256} & {\bf 0.451} & {\bf 0.454} & {\bf 0.337} & {\bf 0.317} & {\bf 0.328} & {\bf 0.447} \\ 
  Lap & 0.542 & 0.311 & 0.283 & 0.306 & 0.541 & 0.524 & 0.400 & 0.381 & 0.399 & 0.526 \\ 
  SAR & 0.537 & 0.316 & 0.289 & 0.312 & 0.522 & 0.531 & 0.415 & 0.395 & 0.414 & 0.529 \\ 
  GP & 0.545 & 0.310 & 0.282 & 0.309 & 0.549 & 0.533 & 0.409 & 0.389 & 0.410 & 0.538 \\ 
  qgam & 3.973 & 2.419 & 1.891 & 2.219 & 3.208 & 1.315 & 1.241 & 1.197 & 1.256 & 1.459 \\ 
  \hline
  &\multicolumn{5}{c|}{MCIW}&\multicolumn{5}{c|}{CP}\\
\hline
 &0.1& 0.3 & 0.5 & 0.7 &0.9&0.1& 0.3 & 0.5 & 0.7 &0.9\\
\hline
  HS & 1.440 & 1.409 & 1.369 & 1.378 & 1.388 & 0.814 & 0.911 & 0.920 & 0.909 & 0.814 \\ 
  Lap & 1.543 & 1.689 & 1.675 & 1.678 & 1.528 & 0.801 & 0.907 & 0.919 & 0.907 & 0.799 \\ 
  SAR & 1.694 & 1.854 & 1.839 & 1.838 & 1.688 & 0.824 & 0.921 & 0.932 & 0.920 & 0.825 \\ 
  GP & 1.681 & 1.864 & 1.858 & 1.852 & 1.662 & 0.823 & 0.924 & 0.937 & 0.924 & 0.823 \\ 
  \hline
  \hline
  &\multicolumn{10}{c|}{Scenario (iii)}\\
\hline
&\multicolumn{5}{c|}{MSE}&\multicolumn{5}{c|}{MAD}\\
\hline
 &0.1& 0.3 & 0.5 & 0.7 &0.9&0.1& 0.3 & 0.5 & 0.7 &0.9\\
\hline
  HS & {\bf 0.651} & {\bf 0.395} & {\bf 0.346} & {\bf 0.375} & {\bf 0.602} & {\bf 0.581} & {\bf 0.438} & {\bf 0.408} & {\bf 0.426} & {\bf 0.559} \\ 
  Lap & 0.740 & 0.431 & 0.389 & 0.424 & 0.740 & 0.643 & 0.493 & 0.469 & 0.489 & 0.642 \\ 
  SAR & 0.736 & 0.427 & 0.387 & 0.422 & 0.721 & 0.648 & 0.501 & 0.479 & 0.499 & 0.644 \\ 
  GP & 0.760 & 0.422 & 0.378 & 0.416 & 0.768 & 0.656 & 0.497 & 0.472 & 0.494 & 0.661 \\ 
  qgam & 3.792 & 2.295 & 1.908 & 2.174 & 2.984 & 1.366 & 1.258 & 1.204 & 1.255 & 1.422 \\ 
  \hline
  &\multicolumn{5}{c|}{MCIW}&\multicolumn{5}{c|}{CP}\\
\hline
 &0.1& 0.3 & 0.5 & 0.7 &0.9&0.1& 0.3 & 0.5 & 0.7 &0.9\\
\hline
  HS & 1.814 & 1.798 & 1.767 & 1.764 & 1.783 & 0.795 & 0.896 & 0.912 & 0.899 & 0.806 \\ 
  Lap & 1.932 & 2.076 & 2.063 & 2.061 & 1.911 & 0.790 & 0.901 & 0.916 & 0.899 & 0.794 \\ 
  SAR & 2.092 & 2.254 & 2.240 & 2.232 & 2.073 & 0.820 & 0.920 & 0.930 & 0.916 & 0.825 \\ 
  GP & 2.071 & 2.276 & 2.273 & 2.270 & 2.048 & 0.815 & 0.925 & 0.938 & 0.925 & 0.821 \\
\hline
\end{tabular}
}
\label{table_1}
\end{center}
\end{table}

\begin{table}[thbp]
\begin{center}
\caption{Average values of MSE, MAD, MCIW, and CP based on $100$ replications for scenarios (iv), (v), and (vi) (exponential function). The minimum values of MSE and MAD are represented in bold.}
\resizebox{1.0\textwidth}{!}{ 
\begin{tabular}{|c|ccccc|ccccc|}
\hline
  &\multicolumn{10}{c|}{Scenario (iv)}\\
\hline
&\multicolumn{5}{c|}{MSE}&\multicolumn{5}{c|}{MAD}\\
\hline
 &0.1& 0.3 & 0.5 & 0.7 &0.9&0.1& 0.3 & 0.5 & 0.7 &0.9\\
\hline
HS & {\bf 0.135} & {\bf 0.082} &  0.075 & 0.082 & {\bf 0.135} & {\bf 0.280} & {\bf 0.218} & {\bf 0.207} & {\bf 0.214} & {\bf 0.274} \\ 
  Lap & 0.184 & {\bf 0.082} & {\bf 0.072} & {\bf 0.081} & 0.186 & 0.341 & 0.224 & 0.212 & 0.226 & 0.342 \\ 
  SAR & 0.195 & 0.087 & 0.076 & 0.085 & 0.184 & 0.357 & 0.234 & 0.220 & 0.233 & 0.347 \\ 
  GP & 0.251 & 0.119 & 0.107 & 0.125 & 0.273 & 0.405 & 0.275 & 0.262 & 0.285 & 0.423 \\ 
  qgam & 0.577 & 0.450 & 0.386 & 0.418 & 0.510 & 0.522 & 0.503 & 0.504 & 0.522 & 0.551 \\
  \hline
  &\multicolumn{5}{c|}{MCIW}&\multicolumn{5}{c|}{CP}\\
\hline
HS & 0.994 & 0.982 & 0.983 & 0.986 & 1.021 & 0.816 & 0.907 & 0.921 & 0.915 & 0.846 \\ 
  Lap & 1.160 & 1.148 & 1.136 & 1.143 & 1.169 & 0.857 & 0.952 & 0.964 & 0.953 & 0.863 \\ 
  SAR & 1.275 & 1.259 & 1.244 & 1.243 & 1.263 & 0.887 & 0.967 & 0.975 & 0.970 & 0.893 \\ 
  GP & 1.333 & 1.424 & 1.433 & 1.445 & 1.348 & 0.878 & 0.961 & 0.972 & 0.961 & 0.872 \\ 
\hline
\hline
  &\multicolumn{10}{c|}{Scenario (v)}\\
\hline
&\multicolumn{5}{c|}{MSE}&\multicolumn{5}{c|}{MAD}\\
\hline
 &0.1& 0.3 & 0.5 & 0.7 &0.9&0.1& 0.3 & 0.5 & 0.7 &0.9\\
\hline
HS & {\bf 0.284} & 0.125 & 0.109 & 0.122 & 0.282 & {\bf 0.355} & {\bf 0.246} & 0.228 & {\bf 0.242} & {\bf 0.349} \\ 
  Lap & 0.313 & 0.116 & 0.095 & 0.115 & 0.312 & 0.397 & 0.248 & {\bf 0.225} & 0.246 & 0.396 \\ 
  SAR & 0.289 & {\bf 0.112} & {\bf 0.093} & {\bf 0.110} & {\bf 0.272} & 0.398 & 0.251 & 0.229 & 0.247 & 0.385 \\ 
  GP & 0.385 & 0.152 & 0.131 & 0.166 & 0.418 & 0.457 & 0.293 & 0.272 & 0.304 & 0.475 \\ 
  qgam & 0.686 & 0.472 & 0.398 & 0.432 & 0.619 & 0.588 & 0.515 & 0.508 & 0.528 & 0.601 \\  
  \hline
  &\multicolumn{5}{c|}{MCIW}&\multicolumn{5}{c|}{CP}\\
\hline
 &0.1& 0.3 & 0.5 & 0.7 &0.9&0.1& 0.3 & 0.5 & 0.7 &0.9\\
\hline
HS & 1.141 & 1.075 & 1.067 & 1.082 & 1.154 & 0.816 & 0.907 & 0.926 & 0.915 & 0.832 \\ 
  Lap & 1.306 & 1.228 & 1.208 & 1.225 & 1.302 & 0.849 & 0.947 & 0.959 & 0.949 & 0.853 \\ 
  SAR & 1.434 & 1.339 & 1.308 & 1.321 & 1.405 & 0.876 & 0.963 & 0.971 & 0.961 & 0.883 \\ 
  GP & 1.514 & 1.533 & 1.525 & 1.557 & 1.517 & 0.858 & 0.959 & 0.969 & 0.955 & 0.853 \\
\hline
\hline
  &\multicolumn{10}{c|}{Scenario (vi)}\\
\hline
&\multicolumn{5}{c|}{MSE}&\multicolumn{5}{c|}{MAD}\\
\hline
 &0.1& 0.3 & 0.5 & 0.7 &0.9&0.1& 0.3 & 0.5 & 0.7 &0.9\\
\hline
HS & {\bf 0.348} & {\bf 0.169} & 0.142 & {\bf 0.151} & {\bf 0.311} & {\bf 0.423} & {\bf 0.305} & {\bf 0.278} & {\bf 0.287} & {\bf 0.400} \\ 
  Lap & 0.435 & 0.170 & 0.140 & 0.157 & 0.408 & 0.487 & 0.307 & 0.280 & 0.296 & 0.470 \\ 
  SAR & 0.424 & {\bf 0.169} & {\bf 0.138} & {\bf 0.151} & 0.362 & 0.495 & 0.313 & 0.283 & 0.295 & 0.456 \\ 
  GP & 0.579 & 0.224 & 0.182 & 0.221 & 0.587 & 0.578 & 0.361 & 0.327 & 0.359 & 0.584 \\ 
  qgam & 0.916 & 0.598 & 0.492 & 0.505 & 0.655 & 0.718 & 0.591 & 0.565 & 0.579 & 0.660 \\
  \hline
  &\multicolumn{5}{c|}{MCIW}&\multicolumn{5}{c|}{CP}\\
\hline
 &0.1& 0.3 & 0.5 & 0.7 &0.9&0.1& 0.3 & 0.5 & 0.7 &0.9\\
\hline
HS & 1.392 & 1.293 & 1.272 & 1.273 & 1.413 & 0.798 & 0.890 & 0.911 & 0.901 & 0.833 \\ 
  Lap & 1.634 & 1.518 & 1.477 & 1.484 & 1.613 & 0.839 & 0.938 & 0.953 & 0.943 & 0.853 \\ 
  SAR & 1.794 & 1.664 & 1.606 & 1.593 & 1.720 & 0.872 & 0.959 & 0.969 & 0.960 & 0.888 \\ 
  GP & 1.887 & 1.886 & 1.853 & 1.894 & 1.872 & 0.849 & 0.957 & 0.971 & 0.960 & 0.860 \\ 
\hline
\end{tabular}
}
\label{table_2}
\end{center}
\end{table}

\section{Application to crime trend analysis in Tokyo}\label{sec:4}

\begin{figure}[htbp]
\begin{center}
\includegraphics[width=\linewidth]{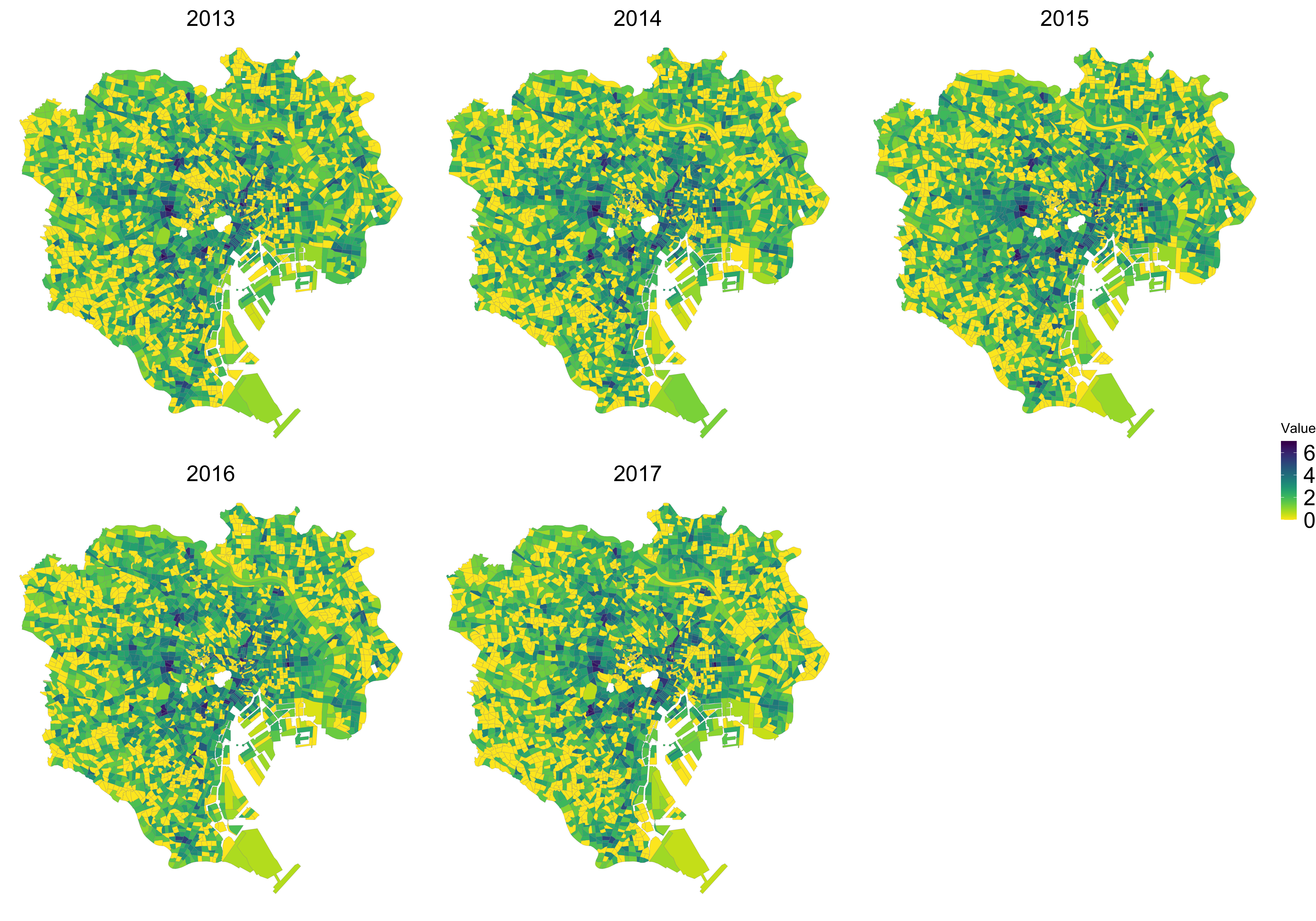}
\caption{Spatial plot of $\log(1+Y)$ for crime density $Y$ based on raw data from 2013 to 2017.}
\label{tokyo_crime_data}
\end{center}
\end{figure}

\begin{figure}[htbp]
\begin{center}
\includegraphics[width=\linewidth]{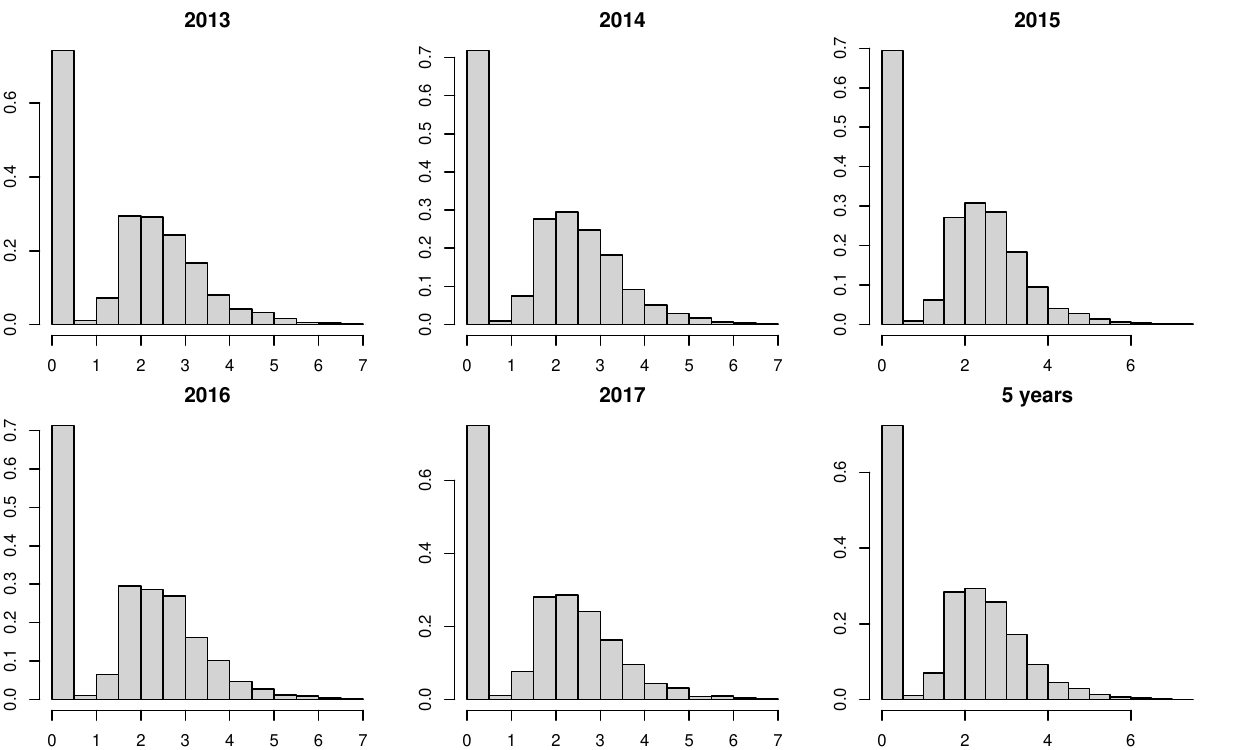}
\caption{The histogram of the crime densities for each year and five-year data.}
\label{tokyo_crime_hist}
\end{center}
\end{figure}

\begin{figure}[htbp]
\begin{center}
\includegraphics[width=12cm]{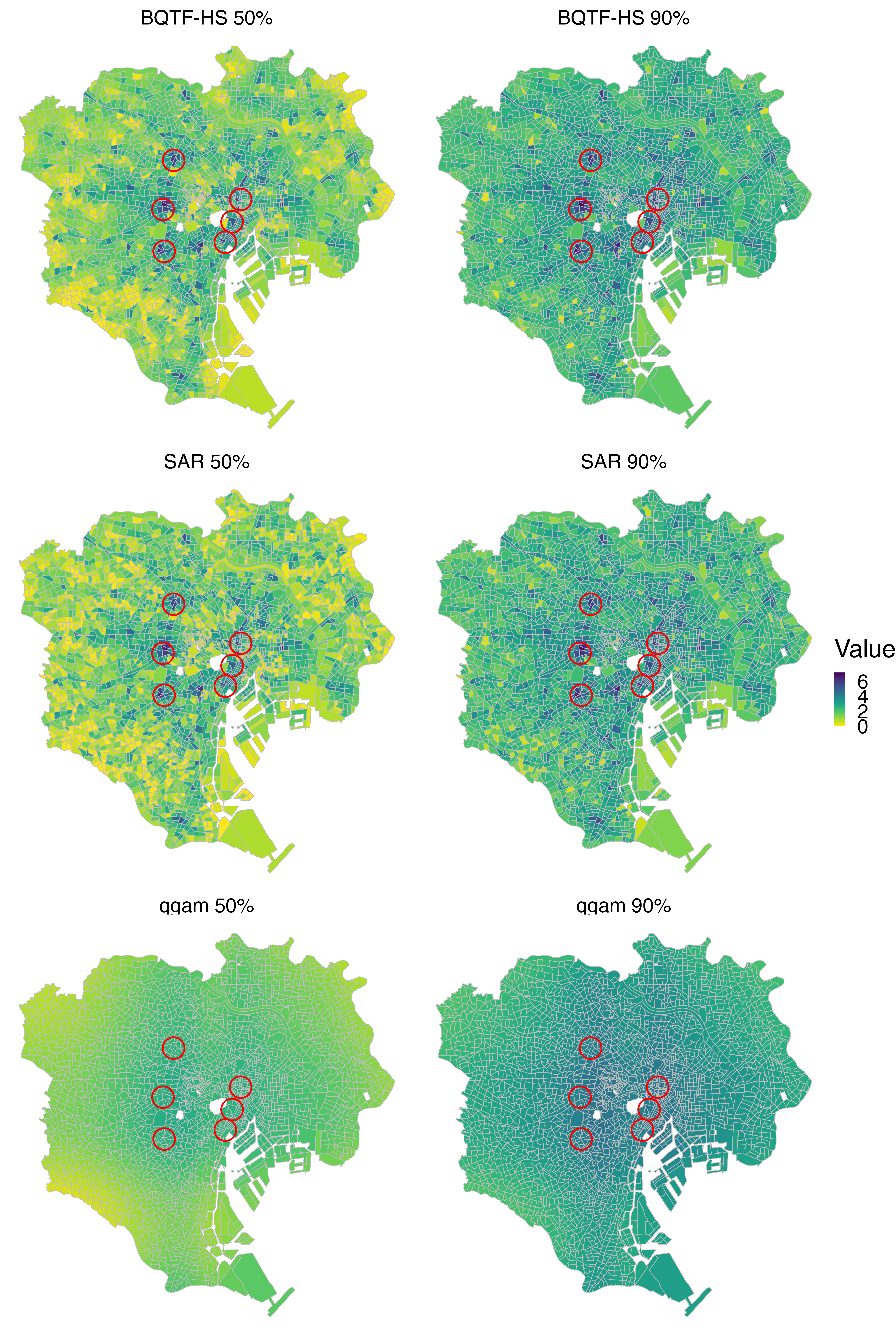}
\caption{Estimated trends via BQTF-HS, SAR, and qgam from top to bottom for two quantile levels: 50\% (left) and 90\% (right). The six red points are the main stations (Shinjuku, Ikebukuro, Shibuya, Shinbashi, Tokyo, and Akihabara) in Tokyo.}
\label{tokyo_crime_estimate}
\end{center}
\end{figure}

We apply the proposed methods to spatial data analysis. We used the ``GIS database of number of police-recorded crime at O-aza, chome in Tokyo, 2009--2017", which was provided by University of Tsukuba Division of Policy and Planning Sciences Commons. The ``chome" represents a specific area, block or street within a city or town. For example, ``3-chome, Shinjuku" would refer to the third block within Shinjuku town in Tokyo. The data contains the number of crimes in Tokyo, and we focus on the violent crime data in particular. We used the number of violent crimes from some 23 wards in Tokyo for five years (from 2013 to 2017) whose number of locations is $n=3,125$ and the sample size is $N=3125\times5=15,625$. The number of edges is 8,996. The edges are constructed based on the 5 nearest neighbor searches. Namely, when $x_j$ is in 5 nearest neighbors of $x_i$, then we connect $x_i$ and $x_j$ even if they are not adjacent to each other on the map. Since the data also involve information on the area (km$^2$) of each region, we define $Y=(Y_1,\dots, Y_{3125})$ as the values of the number of violent crimes divided by the area for each region, which are called {\it crime density} as we mentioned in Section \ref{sec:1}. Using the value of $Y$ may be reasonable because the larger the area, the greater the number of crimes in general. \cite{balocchi2023crime} also used the crime density normalized by the area. Moreover, we use the value on the log scale as data $y=\log(1+Y)$. Such a transformation is popular in the literature \cite[see also][]{balocchi2019spatial}. We regard five years of data as multiple observation data per location. The data is shown in Figure \ref{tokyo_crime_data}, and the plot indicates that spatial trends have not changed over the years. Additionally, the histograms of $y$ for each year and all years are also shown in Figure \ref{tokyo_crime_hist}, which represent that the distribution of observed data is the same for all years. Although there are some hotspots for each year in Figure \ref{tokyo_crime_data}, some high-risk areas have overlapped throughout the five years. They tend to be particularly common in downtown areas, and such areas can be seen as ones with potentially high risk. In this section, our goal is to estimate spatial quantile trends and detect potential hotspots. In particular, since we are interested in the median and high-risk cases of criminal activity, we estimate 50\% and 90\% quantile trends. 
We adopt the proposed Bayesian quantile trend filtering under horseshoe prior (BQTF-HS) and compare the performance with two methods: the SAR model and the additive quantile regression (qgam) using latitude and longitude as covariates. The GP model has a high computation cost because the covariance matrix is not a sparse matrix, unlike the BQTF and SAR models. Therefore, although the GP models can be applied to the example, we only consider the above two methods as competitors.
For the Bayesian methods, we generated 50,000 posterior samples, and then the first 10,000 samples were discarded and only every 40th scan was saved. The order of trend filtering is $k=1$. The estimated quantile trends are shown in Figure \ref{tokyo_crime_estimate}. Note that if the estimate has a negative value, then it is plotted as zero. The proposed BQTF method seems to capture the zero-inflated data throughout five years. For the 90\% quantile trend, the BQTF method provides the adaptive smoothing that detects not only high-risk spots but also low-risk spots and gives smoothing a high quantile trend. 
The estimate of the SAR model is not smoother than the BQTF methods and is similar to raw data shown in Figure \ref{tokyo_crime_data}.
On the other hand, the additive quantile regression (qgam) method results in over-shrinkage and clearly can not achieve a locally adaptive smoothing. In other words, the qgam method cannot detect hotspots, and the areas that seem to be not hotspots also have a green or blue color. 
Therefore, we can conclude that the proposed BQTF method gives a more reasonable estimate of potential quantile trend than the SAR model and qgam, which are not smooth or producing over-shrinkage results. 
The six hotspots detected by the proposed are shown in Figure \ref{tokyo_crime_estimate}, which are the main stations (Shinjuku, Ikebukuro, Shibuya, Shinbashi, Tokyo, and Akihabara) in the Yamanote line, and the areas are filled by blue, which have high values. 
Such areas seem to be outliers in \cite{hamura2021robust}, and the same result is observed. Moreover, the lower left area filled by yellow is considered lower-risk in terms of 50\% trend, and the spatial effect analyzed in \cite{hamura2021robust} also has small values in these areas. However, as seen in the 90\% trend, it seems that the risk in these areas is not low potentially. The potential risk is not clear from the 50\% quantile trend or the other method.
Hence, the BQTF method provides a locally adaptive smoothing for high quantile trends and captures latent heterogeneity by treating five years of data as multiple observations.

\section{Concluding remarks}

In this paper, we proposed a Bayesian quantile trend filtering (BQTF) method on graphs under continuous shrinkage priors, which enables us to estimate quantile trends for spatial data. We also provide a simple Gibbs sampler by introducing a kind of shadow prior. Through simulation studies, it is shown that the BQTF estimates under the horseshoe prior provide locally adaptive smoothing in the sense of capturing the change of quantile trends and estimating the smooth quantile trends. The application of the violent crime data in Tokyo gives interesting results in that the proposed method provides locally adaptive quantile smoothing for all quantiles and detects hotspots focusing on a low quantile level.

There are several future directions for this paper. First, it should be proved some theoretical results for Bayesian quantile trend filtering such as the posterior consistency under misspecified asymmetric Laplace likelihood, the valid uncertainly quantification, and the posterior contraction rate \citep[see also][]{sriram2013posterior, sriram2015sandwich, banerjee2022horseshoe}. Furthermore, since the proposed methods do not work well to estimate extremal quantiles, it is also important to extend the proposed methods to smoothing for extremal quantiles \citep[e.g.][]{chernozhukov2005extremal}. In terms of application, while we use areal data, the trend estimation of point-level data has also been studied \citep[see also][]{lum2012spatial}. Since the observation points of these data are different between years, the proposed methods cannot be used as is to detect potential hotspots throughout multiple years. Finally, although we only modeled spatial smoothing without covariates, it may be important to consider the covariates. Since \cite{sadhanala2019additive} considered an extension of trend filtering to additive models to handle covariates, such an extension of our proposed model will be also expected.

\section*{Acknowledgement}
The authors would like to thank an Associate Editor and two anonymous reviewers for valuable comments and helpful suggestions. 
This work is supported by JST, the establishment of university fellowships towards the creation of science technology innovation, Grant Number JPMJFS2129. This work is partially supported by Japan Society for Promotion of Science (KAKENHI) grant numbers 21K13835 and 21H00699.

\vspace{5mm}
\bibliography{ref}

\begin{thebibliography}{}

\bibitem[\protect\citeauthoryear{Balocchi, Deshpande, George, and
  Jensen}{Balocchi et~al.}{2023}]{balocchi2023crime}
Balocchi, C., S.~K. Deshpande, E.~I. George, and S.~T. Jensen (2023).
\newblock Crime in philadelphia: Bayesian clustering with particle
  optimization.
\newblock {\em Journal of the American Statistical Association\/}~{\em
  118\/}(542), 818--829.

\bibitem[\protect\citeauthoryear{Balocchi and Jensen}{Balocchi and
  Jensen}{2019}]{balocchi2019spatial}
Balocchi, C. and S.~T. Jensen (2019).
\newblock Spatial modeling of trends in crime over time in philadelphia.
\newblock {\em The Annals of Applied Statistics\/}~{\em 13\/}(4), 2235--2259.

\bibitem[\protect\citeauthoryear{Banerjee}{Banerjee}{2022}]{banerjee2022horseshoe}
Banerjee, S. (2022).
\newblock Horseshoe shrinkage methods for bayesian fusion estimation.
\newblock {\em Computational Statistics \& Data Analysis\/}~{\em 174}, 107450.

\bibitem[\protect\citeauthoryear{Barata, Prado, and Sans{\'o}}{Barata
  et~al.}{2022}]{barata2022fast}
Barata, R., R.~Prado, and B.~Sans{\'o} (2022).
\newblock Fast inference for time-varying quantiles via flexible dynamic models
  with application to the characterization of atmospheric rivers.
\newblock {\em The Annals of Applied Statistics\/}~{\em 16\/}(1), 247--271.

\bibitem[\protect\citeauthoryear{Braga}{Braga}{2001}]{braga2001effects}
Braga, A.~A. (2001).
\newblock The effects of hot spots policing on crime.
\newblock {\em The Annals of the American Academy of Political and Social
  Science\/}~{\em 578\/}(1), 104--125.

\bibitem[\protect\citeauthoryear{Brantley, Guinness, and Chi}{Brantley
  et~al.}{2020}]{brantley2020baseline}
Brantley, H.~L., J.~Guinness, and E.~C. Chi (2020).
\newblock Baseline drift estimation for air quality data using quantile trend
  filtering.
\newblock {\em The Annals of Applied Statistics\/}~{\em 14\/}(2), 585--604.

\bibitem[\protect\citeauthoryear{Carvalho, Polson, and Scott}{Carvalho
  et~al.}{2010}]{carvalho2010horseshoe}
Carvalho, C.~M., N.~G. Polson, and J.~G. Scott (2010).
\newblock The horseshoe estimator for sparse signals.
\newblock {\em Biometrika\/}~{\em 97\/}(2), 465--480.

\bibitem[\protect\citeauthoryear{Castillo-Mateo, As{\'\i}n, Cebri{\'a}n,
  Gelfand, and Abaurrea}{Castillo-Mateo et~al.}{2023}]{castillo2023spatial}
Castillo-Mateo, J., J.~As{\'\i}n, A.~C. Cebri{\'a}n, A.~E. Gelfand, and
  J.~Abaurrea (2023).
\newblock Spatial quantile autoregression for season within year daily maximum
  temperature data.
\newblock {\em The Annals of Applied Statistics\/}~{\em 17\/}(3), 2305--2325.

\bibitem[\protect\citeauthoryear{Chernozhukov}{Chernozhukov}{2005}]{chernozhukov2005extremal}
Chernozhukov, V. (2005).
\newblock Extremal quantile regression.
\newblock {\em The Annals of Statistics\/}~{\em 33\/}(2), 806--839.

\bibitem[\protect\citeauthoryear{Fasiolo, Wood, Zaffran, Nedellec, and
  Goude}{Fasiolo et~al.}{2021}]{fasiolo2021fast}
Fasiolo, M., S.~N. Wood, M.~Zaffran, R.~Nedellec, and Y.~Goude (2021).
\newblock Fast calibrated additive quantile regression.
\newblock {\em Journal of the American Statistical Association\/}~{\em
  116\/}(535), 1402--1412.

\bibitem[\protect\citeauthoryear{Faulkner and Minin}{Faulkner and
  Minin}{2018}]{faulkner2018locally}
Faulkner, J.~R. and V.~N. Minin (2018).
\newblock Locally adaptive smoothing with markov random fields and shrinkage
  priors.
\newblock {\em Bayesian analysis\/}~{\em 13\/}(1), 225.

\bibitem[\protect\citeauthoryear{Hamura, Irie, and Sugasawa}{Hamura
  et~al.}{2021}]{hamura2021robust}
Hamura, Y., K.~Irie, and S.~Sugasawa (2021).
\newblock Robust hierarchical modeling of counts under zero-inflation and
  outliers.
\newblock {\em arXiv preprint arXiv:2106.10503\/}.

\bibitem[\protect\citeauthoryear{Heng, Zhou, and Chi}{Heng
  et~al.}{2023}]{heng2023bayesian}
Heng, Q., H.~Zhou, and E.~C. Chi (2023).
\newblock Bayesian trend filtering via proximal markov chain monte carlo.
\newblock {\em Journal of Computational and Graphical Statistics\/}, 1--12.

\bibitem[\protect\citeauthoryear{Kim, Koh, Boyd, and Gorinevsky}{Kim
  et~al.}{2009}]{kim2009ell_1}
Kim, S.-J., K.~Koh, S.~Boyd, and D.~Gorinevsky (2009).
\newblock $\ell_1$ trend filtering.
\newblock {\em SIAM review\/}~{\em 51\/}(2), 339--360.

\bibitem[\protect\citeauthoryear{Kowal, Matteson, and Ruppert}{Kowal
  et~al.}{2019}]{kowal2019dynamic}
Kowal, D.~R., D.~S. Matteson, and D.~Ruppert (2019).
\newblock Dynamic shrinkage processes.
\newblock {\em Journal of the Royal Statistical Society: Series B (Statistical
  Methodology)\/}~{\em 81\/}(4), 781--804.

\bibitem[\protect\citeauthoryear{Kozumi and Kobayashi}{Kozumi and
  Kobayashi}{2011}]{kozumi2011gibbs}
Kozumi, H. and G.~Kobayashi (2011).
\newblock Gibbs sampling methods for bayesian quantile regression.
\newblock {\em Journal of Statistical Computation and Simulation\/}~{\em
  81\/}(11), 1565--1578.

\bibitem[\protect\citeauthoryear{Liechty, Liechty, and M{\"u}ller}{Liechty
  et~al.}{2009}]{liechty2009shadow}
Liechty, M.~W., J.~C. Liechty, and P.~M{\"u}ller (2009).
\newblock The shadow prior.
\newblock {\em Journal of Computational and Graphical Statistics\/}~{\em
  18\/}(2), 368--383.

\bibitem[\protect\citeauthoryear{Liu, Chakraborty, Li, Liu, and Lozano}{Liu
  et~al.}{2014}]{liu2014bayesian}
Liu, F., S.~Chakraborty, F.~Li, Y.~Liu, and A.~C. Lozano (2014).
\newblock Bayesian regularization via graph laplacian.
\newblock {\em Bayesian Analysis\/}~{\em 9\/}(2), 449--474.

\bibitem[\protect\citeauthoryear{Lum and Gelfand}{Lum and
  Gelfand}{2012}]{lum2012spatial}
Lum, K. and A.~E. Gelfand (2012).
\newblock Spatial quantile multiple regression using the asymmetric laplace
  process.
\newblock {\em Bayesian Analysis\/}~{\em 7\/}(2), 235--258.

\bibitem[\protect\citeauthoryear{Onizuka, Hashimoto, and Sugasawa}{Onizuka
  et~al.}{2022}]{onizuka2022fast}
Onizuka, T., S.~Hashimoto, and S.~Sugasawa (2022).
\newblock Fast and locally adaptive bayesian quantile smoothing using
  calibrated variational approximations.
\newblock {\em arXiv preprint arXiv:2211.04666\/}.

\bibitem[\protect\citeauthoryear{Park and Casella}{Park and
  Casella}{2008}]{park2008bayesian}
Park, T. and G.~Casella (2008).
\newblock The bayesian lasso.
\newblock {\em Journal of the American Statistical Association\/}~{\em
  103\/}(482), 681--686.

\bibitem[\protect\citeauthoryear{Polson and Scott}{Polson and
  Scott}{2012}]{polson2012half}
Polson, N.~G. and J.~G. Scott (2012).
\newblock On the half-cauchy prior for a global scale parameter.
\newblock {\em Bayesian Analysis\/}~{\em 7\/}(4), 887 -- 902.

\bibitem[\protect\citeauthoryear{Ramdas and Tibshirani}{Ramdas and
  Tibshirani}{2016}]{ramdas2016fast}
Ramdas, A. and R.~J. Tibshirani (2016).
\newblock Fast and flexible admm algorithms for trend filtering.
\newblock {\em Journal of Computational and Graphical Statistics\/}~{\em
  25\/}(3), 839--858.

\bibitem[\protect\citeauthoryear{Reich, Fuentes, and Dunson}{Reich
  et~al.}{2011}]{reich2011bayesian}
Reich, B.~J., M.~Fuentes, and D.~B. Dunson (2011).
\newblock Bayesian spatial quantile regression.
\newblock {\em Journal of the American Statistical Association\/}~{\em
  106\/}(493), 6--20.

\bibitem[\protect\citeauthoryear{Roualdes}{Roualdes}{2015}]{roualdes2015bayesian}
Roualdes, E.~A. (2015).
\newblock Bayesian trend filtering.
\newblock {\em arXiv preprint arXiv:1505.07710\/}.

\bibitem[\protect\citeauthoryear{Sadhanala and Tibshirani}{Sadhanala and
  Tibshirani}{2019}]{sadhanala2019additive}
Sadhanala, V. and R.~J. Tibshirani (2019).
\newblock Additive models with trend filtering.
\newblock {\em The Annals of Statistics\/}~{\em 47\/}(6), 3032--3068.

\bibitem[\protect\citeauthoryear{Sriram}{Sriram}{2015}]{sriram2015sandwich}
Sriram, K. (2015).
\newblock A sandwich likelihood correction for bayesian quantile regression
  based on the misspecified asymmetric laplace density.
\newblock {\em Statistics \& Probability Letters\/}~{\em 107}, 18--26.

\bibitem[\protect\citeauthoryear{Sriram, Ramamoorthi, and Ghosh}{Sriram
  et~al.}{2013}]{sriram2013posterior}
Sriram, K., R.~Ramamoorthi, and P.~Ghosh (2013).
\newblock Posterior consistency of bayesian quantile regression based on the
  misspecified asymmetric laplace density.
\newblock {\em Bayesian Analysis\/}~{\em 8\/}(2), 479--504.

\bibitem[\protect\citeauthoryear{Taddy}{Taddy}{2010}]{taddy2010autoregressive}
Taddy, M.~A. (2010).
\newblock Autoregressive mixture models for dynamic spatial poisson processes:
  Application to tracking intensity of violent crime.
\newblock {\em Journal of the American Statistical Association\/}~{\em
  105\/}(492), 1403--1417.

\bibitem[\protect\citeauthoryear{Tibshirani}{Tibshirani}{2014}]{tibshirani2014adaptive}
Tibshirani, R.~J. (2014).
\newblock Adaptive piecewise polynomial estimation via trend filtering.
\newblock {\em The Annals of statistics\/}~{\em 42\/}(1), 285--323.

\bibitem[\protect\citeauthoryear{Tibshirani and Taylor}{Tibshirani and
  Taylor}{2011}]{tibshirani2011solution}
Tibshirani, R.~J. and J.~Taylor (2011).
\newblock The solution path of the generalized lasso.
\newblock {\em The Annals of Statistics\/}~{\em 39\/}(3), 1335--1371.

\bibitem[\protect\citeauthoryear{Wakayama and Sugasawa}{Wakayama and
  Sugasawa}{2023}]{wakayama2023trend}
Wakayama, T. and S.~Sugasawa (2023).
\newblock Trend filtering for functional data.
\newblock {\em Stat\/}~{\em 12\/}(1), e590.

\bibitem[\protect\citeauthoryear{Wang, Sharpnack, Smola, and Tibshirani}{Wang
  et~al.}{2015}]{wang2015trend}
Wang, Y.-X., J.~Sharpnack, A.~Smola, and R.~Tibshirani (2015).
\newblock Trend filtering on graphs.
\newblock In {\em Artificial Intelligence and Statistics}, pp.\  1042--1050.
  PMLR.

\bibitem[\protect\citeauthoryear{Xu and Ghosh}{Xu and
  Ghosh}{2015}]{xu2015bayesian}
Xu, X. and M.~Ghosh (2015).
\newblock Bayesian variable selection and estimation for group lasso.
\newblock {\em Bayesian Analysis\/}~{\em 10\/}(4), 909--936.

\bibitem[\protect\citeauthoryear{Yano, Kaneko, and Komaki}{Yano
  et~al.}{2021}]{yano2021minimax}
Yano, K., R.~Kaneko, and F.~Komaki (2021).
\newblock Minimax predictive density for sparse count data.
\newblock {\em Bernoulli\/}~{\em 27\/}(2), 1212--1238.

\bibitem[\protect\citeauthoryear{Yu and Moyeed}{Yu and
  Moyeed}{2001}]{yu2001bayesian}
Yu, K. and R.~A. Moyeed (2001).
\newblock Bayesian quantile regression.
\newblock {\em Statistics \& Probability Letters\/}~{\em 54\/}(4), 437--447.

\end{thebibliography}
\bibliographystyle{chicago}

\newpage
\setcounter{page}{1}
\setcounter{equation}{0}
\renewcommand{\theequation}{S\arabic{equation}}
\setcounter{section}{0}
\renewcommand{\thesection}{S\arabic{section}}
\setcounter{table}{0}
\renewcommand{\thetable}{S\arabic{table}}
\setcounter{figure}{0}
\renewcommand{\thefigure}{S\arabic{figure}}

\begin{center}
{\LARGE\bf Supplementary Materials for ``Locally Adaptive Spatial Quantile Smoothing: Application to Monitoring Crime Density in Tokyo"}
\end{center}




\vspace{1cm}
This Supplementary Material provides additional information for the simulation study, the MCMC algorithm of the other methods, and the additional analysis for Tokyo Crime data.

\section{Additional information for simulation study}

\subsection{The MCMC algorithms of SAR and GP models}

In this subsection, we summarized the MCMC algorithms of SAR and GP models compared with the proposed methods in simulation studies. Since the prior of $\theta$ is $\theta\sim N_n(0,\sigma^2\tau^2Q(\rho))$ for both methods, the algorithm of SAR and GP prior is equal. Because $\theta\sim N_n(0,\sigma^2\tau^2Q(\rho))$ is assumed instead of $\theta\mid \sigma^2,\tau^2, w\sim N_n(0, \sigma^2\tau^2(D^{\top}W^{-1}D)^{-1})$ in the proposed methods, the algorithm is directly given as follows:
\begin{itemize}
\item Sample $\theta$ from
\begin{align*}
    &\theta\mid y, \sigma^2, z, \tau^2, \rho \sim N(A^{-1}B, \sigma^2A^{-1}),\\
    &A=\frac{1}{\tau^2}Q^{-1}+\frac{1}{t^2}\diag\left(\sum_{j=1}^{N_1}z_{1j}^{-1},\ldots,\sum_{j=1}^{N_n}z_{nj}^{-1}\right), \\
    &B=\left(\sum_{j=1}^{N_1}\frac{y_{1j}-\psi z_{1j}}{t^2z_{1j}},\dots,\sum_{j=1}^{N_n}\frac{y_{nj}-\psi z_{nj}}{t^2z_{nj}}\right)^{\top},\quad Q=Q(\rho).
\end{align*}
\item Sample $\sigma^2$ from 
\begin{align*}
&\sigma^2\mid \theta, y, z \sim \mathrm{IG}\left(a_{\sigma}+\frac{n+3N}{2}, \beta_{\sigma}\right),\\
&\beta_{\sigma}=b_{\sigma}+\sum_{i=1}^n\sum_{j=1}^{N_i}\frac{(y_{ij}-\theta_i-\psi z_{ij})^2}{2t^2z_{ij}}+\sum_{i=1}^n\sum_{j=1}^{N_i} z_{ij}+\frac{1}{\tau^2}\theta^{\top}Q^{-1}\theta.
\end{align*}
\item If $\tau^2\sim \mathrm{IG}(a_{\tau}, b_{\tau})$ is assumed for the prior of $\tau^2$, sample $\tau^2$ from
\begin{align*}
\tau^2\mid \theta,\sigma^2 \sim \mathrm{IG}\left(a_{\tau}+n/2, b_{\tau}+\frac{1}{2\sigma^2}\theta^{\top}Q^{-1}\theta\right).
\end{align*}
\item The parameter $\rho$ is sampled with the random walk MH. 
\end{itemize}

\subsection{True five quantiles under the three noise distribution}

\begin{figure}[htbp]
\begin{center}
\includegraphics[width=\linewidth]{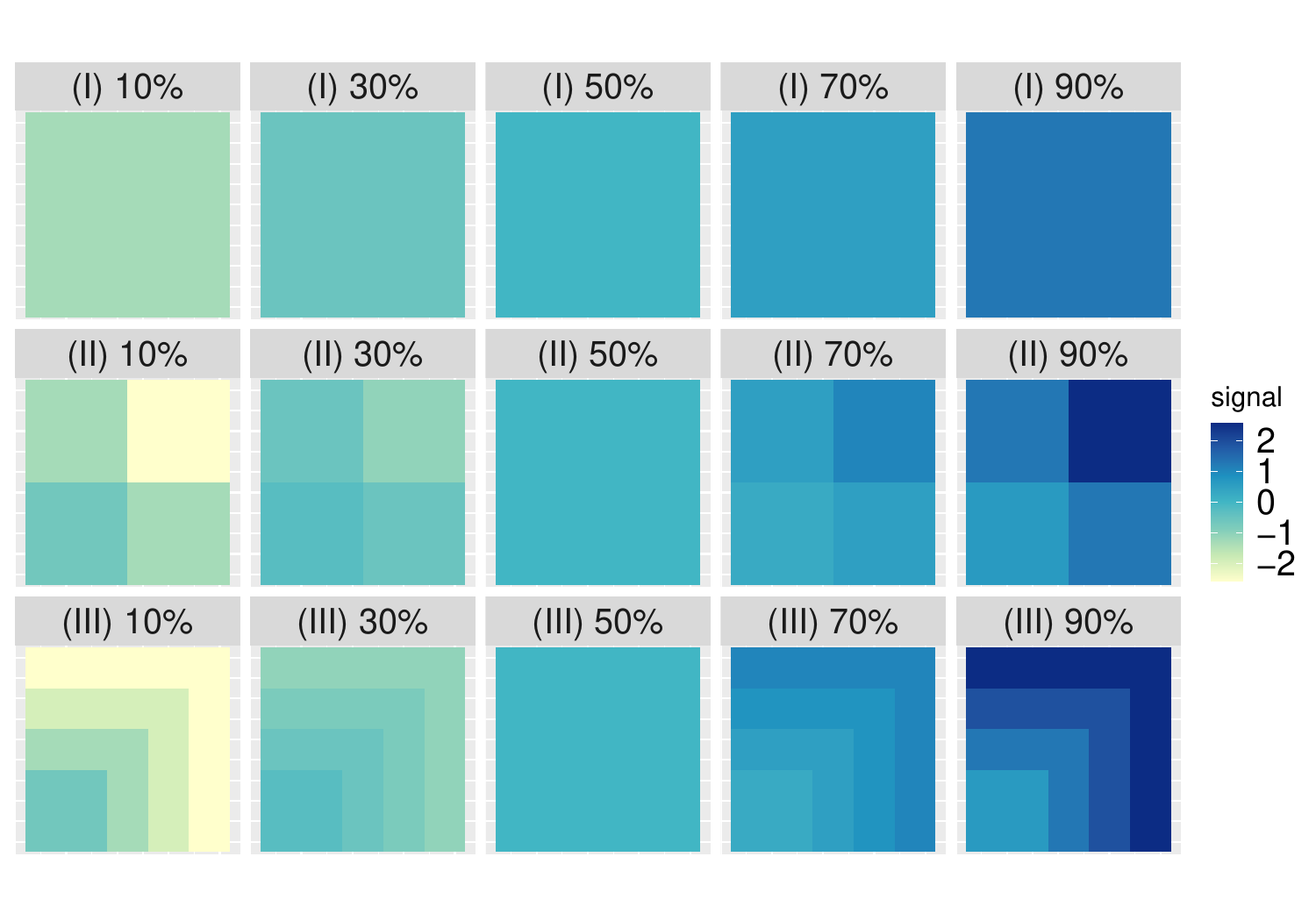}
\caption{The true $p$-th quantiles ($p=0.1,0.3,0.5,0.7,0.9$) for three noise distributions from top to bottom: (I), (II), (III).}
\label{Sim_noise}
\end{center}
\end{figure}

Figure \ref{Sim_noise} summarizes the plot of true five quantile trends under three noise distribution $\epsilon$ in the simulation study (see also Section 3). We considered the three scenarios: that is (I) homogeneous, (II) block heterogeneous, and (III) smooth heterogeneous.

\subsection{Computation time and efficiency}

We provide the raw computing time and the sampling efficiency of the MCMC algorithm in the simulation study. We calculated the effective sample size per unit time, defined as the effective sample size divided by the computation time in seconds. The results are reported in Tables~\ref{table_ESSTime1} and \ref{table_ESSTime2}. Although the effective sample size (ESS) per computation time under Laplace prior is larger than the other methods, the computation times are similar. It is also observed that there are no differences between scenarios.

\begin{table}[thbp]
\caption{Average values of effective sample size per unit time and raw computing time based on $100$ replications for scenarios (i), (ii), and (iii). }
\begin{center}
\begin{tabular}{|c|ccccc|ccccc|}
\hline
&\multicolumn{10}{c|}{Scenario (i)}\\
\hline
&\multicolumn{5}{c|}{ESS (per second)}&\multicolumn{5}{c|}{Compuation rime (second)}\\
\hline
 &0.1& 0.3 & 0.5 & 0.7 &0.9&0.1& 0.3 & 0.5 & 0.7 &0.9\\
\hline
  HS & 11 & 16 & 19 & 17 & 11 & 66 & 66 & 66 & 66 & 66 \\ 
  Lap & 20 & 34 & 38 & 35 & 20 & 71 & 70 & 70 & 71 & 71 \\ 
  SAR & 11 & 12 & 12 & 12 & 11 & 64 & 64 & 64 & 64 & 64 \\ 
  GP & 11 & 12 & 12 & 12 & 11 & 66 & 66 & 66 & 66 & 66 \\  
\hline
\hline
&\multicolumn{10}{c|}{Scenario (ii)}\\
\hline
&\multicolumn{5}{c|}{ESS (per second)}&\multicolumn{5}{c|}{Compuation rime (second)}\\
\hline
 &0.1& 0.3 & 0.5 & 0.7 &0.9&0.1& 0.3 & 0.5 & 0.7 &0.9\\
\hline
  HS & 11 & 17 & 20 & 18 & 11 & 66 & 66 & 66 & 66 & 66 \\  
  Lap & 20 & 36 & 40 & 36 & 20 & 71 & 70 & 70 & 70 & 71 \\ 
  SAR & 11 & 12 & 12 & 12 & 11 & 64 & 64 & 64 & 64 & 64 \\ 
  GP & 11 & 12 & 12 & 12 & 11 & 66 & 66 & 66 & 66 & 66 \\ 
  \hline
  \hline
  &\multicolumn{10}{c|}{Scenario (iii)}\\
\hline
&\multicolumn{5}{c|}{ESS (per second)}&\multicolumn{5}{c|}{Compuation rime (second)}\\
\hline
 &0.1& 0.3 & 0.5 & 0.7 &0.9&0.1& 0.3 & 0.5 & 0.7 &0.9\\
\hline
  HS & 10 & 14 & 16 & 15 & 10 & 66 & 66 & 66 & 66 & 66 \\  
  Lap & 21 & 35 & 39 & 36 & 21 & 70 & 70 & 70 & 70 & 70 \\ 
  SAR & 11 & 12 & 12 & 12 & 11 & 64 & 64 & 64 & 64 & 64 \\ 
  GP & 11 & 12 & 12 & 12 & 11 & 66 & 66 & 66 & 66 & 66 \\  
\hline
\end{tabular}
\end{center}
\label{table_ESSTime1}
\end{table}

\begin{table}[thbp]
\caption{Average values of effective sample size and raw computing time based on $100$ replications for scenarios (iv), (v), and (vi). }
\begin{center}
\begin{tabular}{|c|ccccc|ccccc|}
\hline
&\multicolumn{10}{c|}{Scenario (iv)}\\
\hline
&\multicolumn{5}{c|}{ESS (per second)}&\multicolumn{5}{c|}{Compuation rime (second)}\\
\hline
 &0.1& 0.3 & 0.5 & 0.7 &0.9&0.1& 0.3 & 0.5 & 0.7 &0.9\\
\hline
HS & 10 & 11 & 11 & 11 & 10 & 65 & 65 & 65 & 65 & 65 \\ 
  Lap & 21 & 35 & 37 & 35 & 21 & 70 & 69 & 69 & 69 & 70 \\ 
  SAR & 11 & 12 & 12 & 12 & 11 & 64 & 64 & 64 & 64 & 64 \\ 
  GP & 11 & 11 & 11 & 11 & 11 & 67 & 67 & 67 & 67 & 67 \\ 
\hline
\hline
&\multicolumn{10}{c|}{Scenario (v)}\\
\hline
&\multicolumn{5}{c|}{ESS (per second)}&\multicolumn{5}{c|}{Compuation rime (second)}\\
\hline
 &0.1& 0.3 & 0.5 & 0.7 &0.9&0.1& 0.3 & 0.5 & 0.7 &0.9\\
\hline
HS & 10 & 11 & 11 & 11 & 9 & 65 & 65 & 65 & 65 & 65 \\ 
  Lap & 22 & 36 & 39 & 37 & 21 & 70 & 69 & 69 & 69 & 70 \\ 
  SAR & 11 & 12 & 12 & 12 & 11 & 64 & 64 & 64 & 64 & 64 \\ 
  GP & 11 & 11 & 11 & 11 & 11 & 67 & 67 & 67 & 67 & 67 \\ 
  \hline
  \hline
  &\multicolumn{10}{c|}{Scenario (vi)}\\
\hline
&\multicolumn{5}{c|}{ESS (per second)}&\multicolumn{5}{c|}{Compuation rime (second)}\\
\hline
 &0.1& 0.3 & 0.5 & 0.7 &0.9&0.1& 0.3 & 0.5 & 0.7 &0.9\\
\hline
HS & 9 & 11 & 11 & 11 & 9 & 66 & 66 & 66 & 66 & 66 \\ 
  Lap & 22 & 36 & 39 & 36 & 22 & 70 & 70 & 70 & 70 & 70 \\ 
  SAR & 11 & 12 & 12 & 12 & 11 & 64 & 64 & 64 & 64 & 64 \\ 
  GP & 11 & 11 & 11 & 11 & 11 & 67 & 67 & 67 & 67 & 67 \\
\hline
\end{tabular}
\end{center}
\label{table_ESSTime2}
\end{table}

\subsection{Comparison with the other methods}

In addition to the methods presented in Section 3, we compared the following methods. 
\begin{itemize}
\item HS ($k=0$): the proposed method under horseshoe prior and 1st order difference operator. 
\item Lap ($k=0$): the proposed method under Laplace prior and 1st order difference operator. 
\item spreg: the classical spatial regression with two covariates (two-dimensional coordinate) and spatially correlated error terms, which is implemented by using the package {\tt spatialreg} in {\tt R}. Since we can not estimate a quantile using the method, we only compare it with the 50\% quantile trend (true signal).
\end{itemize}
The results are reported in Tables~\ref{table_S1} and \ref{table_S2}. We also showed the HS and Lap methods under $k=1$ for comparison. For the lower quantile level, because the BQTF methods under $k=0$ lead to strong shrinkage, the point estimates of $k=0$ are worse than those of $k=0$ in the 0.1-th quantile level, especially under Laplace prior. The estimates under $k=0$ are better than those of $k=0$ for the other quantile levels. As seen in the main manuscript, the results of $k=1$ under the two-block structure are better than the existing methods, and then we adopt $k=1$ for real data analysis. Compared with spreg in Table~\ref{table_2}, it is observed that the method does not work as well as the qgam method presented in the main manuscript.

\begin{table}[thbp]
\caption{Average values of MSE, MAD, MCIW, and CP based on $100$ replications for scenarios (i), (ii), and (iii) (two-block structure).}
\begin{center}
\resizebox{1.0\textwidth}{!}{ 
\begin{tabular}{|c|ccccc|ccccc|}
\hline
&\multicolumn{10}{c|}{Scenario (i)}\\
\hline
&\multicolumn{5}{c|}{MSE}&\multicolumn{5}{c|}{MAD}\\
\hline
 &0.1& 0.3 & 0.5 & 0.7 &0.9&0.1& 0.3 & 0.5 & 0.7 &0.9\\
\hline
HS ($k=0$) & 0.545 & 0.052 & 0.046 & 0.053 & 0.106 & 0.329 & 0.174 & 0.165 & 0.176 & 0.247 \\ 
  HS ($k=1$) & 0.264 & 0.158 & 0.138 & 0.152 & 0.248 & 0.377 & 0.283 & 0.266 & 0.280 & 0.371 \\ 
  Lap ($k=0$) & 5.649 & 0.128 & 0.115 & 0.125 & 0.221 & 1.283 & 0.277 & 0.264 & 0.276 & 0.373 \\ 
  Lap ($k=1$) & 0.337 & 0.212 & 0.193 & 0.211 & 0.339 & 0.460 & 0.359 & 0.344 & 0.360 & 0.463 \\ 
  \hline
  &\multicolumn{5}{c|}{MCIW}&\multicolumn{5}{c|}{CP}\\
\hline
 & 0.1 & 0.3 & 0.5 & 0.7 & 0.9 & 0.1 & 0.3 & 0.5 & 0.7 & 0.9 \\ 
  \hline
HS ($k=0$) & 0.957 & 1.011 & 1.015 & 1.012 & 1.027 & 0.857 & 0.975 & 0.983 & 0.978 & 0.914 \\ 
  HS ($k=1$) & 1.322 & 1.287 & 1.262 & 1.261 & 1.290 & 0.822 & 0.920 & 0.934 & 0.921 & 0.827 \\ 
  Lap ($k=0$) & 0.228 & 1.363 & 1.355 & 1.353 & 1.318 & 0.247 & 0.947 & 0.955 & 0.945 & 0.876 \\ 
  Lap ($k=1$)  & 1.400 & 1.554 & 1.552 & 1.541 & 1.395 & 0.799 & 0.910 & 0.921 & 0.906 & 0.801 \\ 
\hline
\hline
&\multicolumn{10}{c|}{Scenario (ii)}\\
\hline
&\multicolumn{5}{c|}{MSE}&\multicolumn{5}{c|}{MAD}\\
\hline
 &0.1& 0.3 & 0.5 & 0.7 &0.9&0.1& 0.3 & 0.5 & 0.7 &0.9\\
\hline
HS ($k=0$) & 0.319 & 0.117 & 0.088 & 0.119 & 0.328 & 0.368 & 0.223 & 0.190 & 0.224 & 0.369 \\ 
  HS ($k=1$) & 0.464 & 0.272 & 0.244 & 0.256 & 0.451 & 0.454 & 0.337 & 0.317 & 0.328 & 0.447 \\ 
  Lap ($k=0$) & 5.946 & 0.189 & 0.168 & 0.188 & 0.572 & 1.546 & 0.311 & 0.290 & 0.308 & 0.483 \\ 
  Lap ($k=1$) & 0.542 & 0.311 & 0.283 & 0.306 & 0.541 & 0.524 & 0.400 & 0.381 & 0.399 & 0.526 \\ 
  \hline
  &\multicolumn{5}{c|}{MCIW}&\multicolumn{5}{c|}{CP}\\
\hline
 &0.1& 0.3 & 0.5 & 0.7 &0.9&0.1& 0.3 & 0.5 & 0.7 &0.9\\
\hline
HS ($k=0$)  & 1.186 & 1.139 & 1.116 & 1.140 & 1.176 & 0.842 & 0.950 & 0.976 & 0.949 & 0.847 \\ 
  HS ($k=1$) & 1.440 & 1.409 & 1.369 & 1.378 & 1.388 & 0.814 & 0.911 & 0.920 & 0.909 & 0.814 \\ 
  Lap ($k=0$)  & 0.294 & 1.474 & 1.448 & 1.460 & 1.446 & 0.261 & 0.939 & 0.949 & 0.931 & 0.841 \\ 
  Lap ($k=1$)& 1.543 & 1.689 & 1.675 & 1.678 & 1.528 & 0.801 & 0.907 & 0.919 & 0.907 & 0.799 \\ 
  \hline
  \hline
  &\multicolumn{10}{c|}{Scenario (iii)}\\
\hline
&\multicolumn{5}{c|}{MSE}&\multicolumn{5}{c|}{MAD}\\
\hline
 &0.1& 0.3 & 0.5 & 0.7 &0.9&0.1& 0.3 & 0.5 & 0.7 &0.9\\
\hline
HS ($k=0$)  & 0.347 & 0.129 & 0.101 & 0.129 & 0.349 & 0.422 & 0.257 & 0.227 & 0.255 & 0.416 \\ 
  HS ($k=1$) & 0.651 & 0.395 & 0.346 & 0.375 & 0.602 & 0.581 & 0.438 & 0.408 & 0.426 & 0.559 \\ 
  Lap ($k=0$)  & 6.774 & 0.265 & 0.230 & 0.249 & 1.232 & 1.780 & 0.386 & 0.362 & 0.376 & 0.708 \\ 
  Lap ($k=1$) & 0.740 & 0.431 & 0.389 & 0.424 & 0.740 & 0.643 & 0.493 & 0.469 & 0.489 & 0.642 \\ 
  \hline
  &\multicolumn{5}{c|}{MCIW}&\multicolumn{5}{c|}{CP}\\
\hline
 &0.1& 0.3 & 0.5 & 0.7 &0.9&0.1& 0.3 & 0.5 & 0.7 &0.9\\
\hline
HS ($k=0$)  & 1.478 & 1.398 & 1.373 & 1.401 & 1.477 & 0.848 & 0.960 & 0.974 & 0.962 & 0.845 \\ 
  HS ($k=1$)  & 1.814 & 1.798 & 1.767 & 1.764 & 1.783 & 0.795 & 0.896 & 0.912 & 0.899 & 0.806 \\ 
  Lap ($k=0$)  & 0.382 & 1.767 & 1.726 & 1.737 & 1.670 & 0.059 & 0.929 & 0.940 & 0.927 & 0.784 \\ 
  Lap ($k=1$) & 1.932 & 2.076 & 2.063 & 2.061 & 1.911 & 0.790 & 0.901 & 0.916 & 0.899 & 0.794 \\  
\hline
\end{tabular}
}
\end{center}
\label{table_S1}
\end{table}

\begin{table}[thbp]
\caption{Average values of MSE, MAD, MCIW, and CP based on $100$ replications for all scenarios. }
\begin{center}
\begin{tabular}{|c|cccccc|}
\hline
&\multicolumn{6}{c|}{MSE}\\
\hline
 &(i)& (ii) & (iii) & (iv) & (v) & (vi) \\
\hline
HS & 0.138 & 0.244 & 0.346 & 0.075 & 0.109 & 0.142 \\ 
  Lap & 0.193 & 0.283 & 0.389 & 0.072 & 0.095 & 0.140 \\ 
  spreg & 1.744 & 4.209 & 5.886 & 1.791 & 2.855 & 4.523 \\ 
\hline
&\multicolumn{6}{c|}{MAD}\\
\hline
\hline
HS & 0.266 & 0.317 & 0.408 & 0.207 & 0.228 & 0.278 \\ 
  Lap & 0.344 & 0.381 & 0.469 & 0.212 & 0.225 & 0.280 \\ 
  spreg & 0.926 & 1.783 & 2.140 & 1.239 & 1.455 & 1.859 \\ 
\hline
\end{tabular}
\end{center}
\label{table_S2}
\end{table}

\subsection{2-D random graph}

\begin{figure}[htbp]
\begin{center}
\includegraphics[width=12cm]{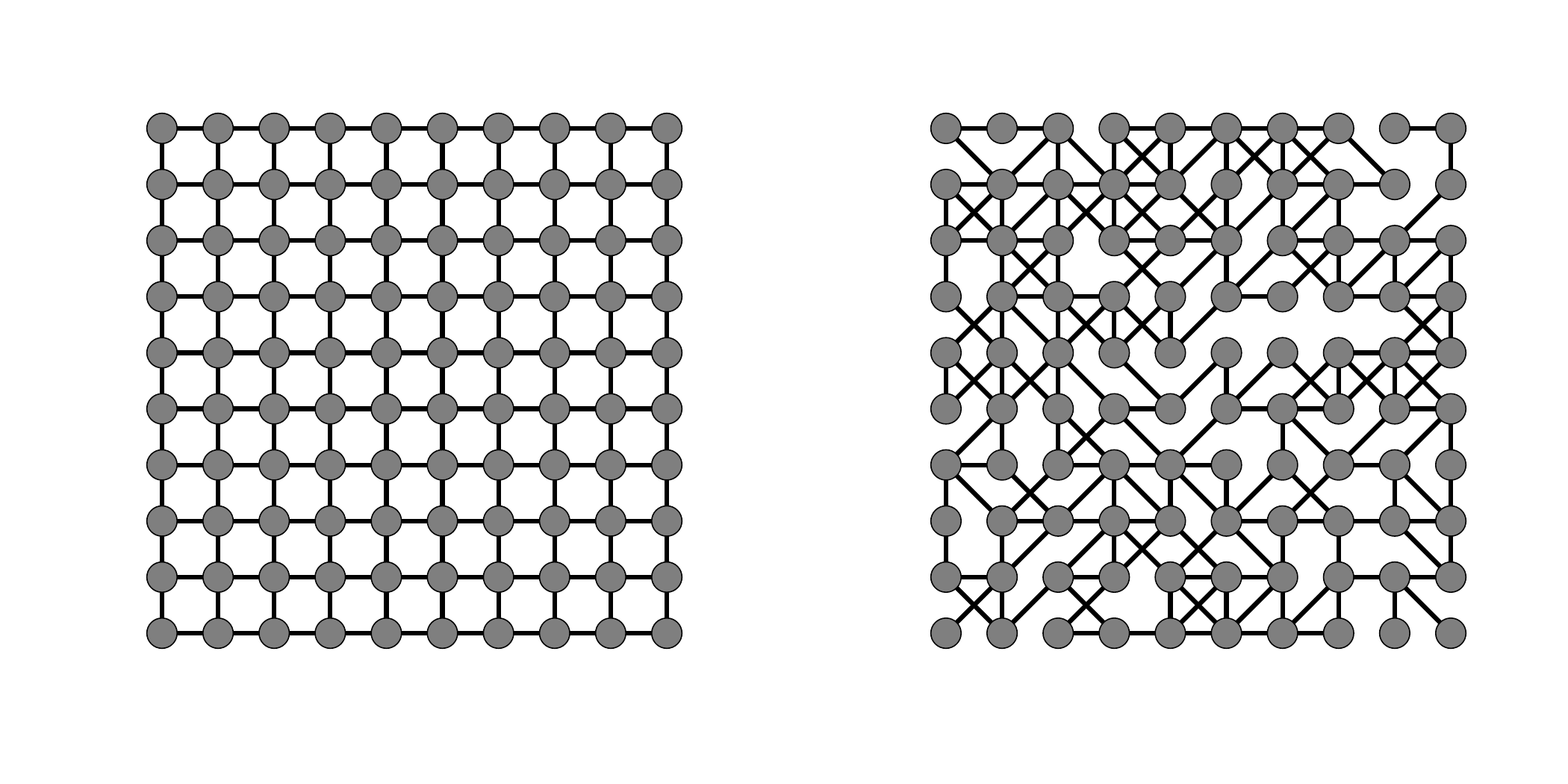}
\caption{The true structures of the 2-D lattice graph (left) and the 2-D random graph (right).}
\label{2Drandom}
\end{center}
\end{figure}

We consider a more general graph based on the 2-D lattice graph in the main simulation study. We set a new 2-D graph with an additional edge drawn on the diagonal, in which the number of vertexes and edges are $|V|=100$ and $|E|=342$, and then the twenty hundred edges are selected from the edge set randomly. The true structures of the 2-D lattice graph and the 2-D random graph are shown in Figure~\ref{2Drandom}. On the graph structure, the simulation studies based on two true signals (two-block structure and exponential function) and three noise distributions ((I), (II), and (III)) are set as the additional simulation study. 
The results are shown in Table~\ref{table_random_1} and \ref{table_random_2}. Note that the results of the compared GP and qgam methods are the same as the 2-D lattice graph because they are only based on the location of the area, not the graph structure. For the two-block structure, the MSE and MAD of the BQTF-HS are the smallest in almost all cases. For the exponential function ((iv), (v), and (vi)), the SAR and the GP models are sometimes better than the proposed methods in center quantiles such as 0.3, 0.5, and 0.7 due to smooth trend. 
Note that the estimates under the 2-D random graph are worse than those under the 2-D lattice graph because the graph is not straightforward, unlike the GP model.

\begin{table}[thbp]
\caption{Average values of MSE, MAD, MCIW, and CP based on $100$ replications for scenarios (i), (ii), and (iii) (two-block structure). The minimum values of MSE and MAD are represented in bold. }
\begin{center}
\resizebox{1.0\textwidth}{!}{ 
\begin{tabular}{|c|ccccc|ccccc|}
\hline
&\multicolumn{10}{c|}{Scenario (i)}\\
\hline
&\multicolumn{5}{c|}{MSE}&\multicolumn{5}{c|}{MAD}\\
\hline
 &0.1& 0.3 & 0.5 & 0.7 &0.9&0.1& 0.3 & 0.5 & 0.7 &0.9\\
\hline 
  HS & {\bf 0.300} & {\bf 0.199} & {\bf 0.178} & {\bf 0.191} & {\bf 0.290} & {\bf 0.415} & {\bf 0.331} & {\bf 0.314} & {\bf 0.327} & {\bf 0.409} \\ 
  Lap & 0.346 & 0.214 & 0.196 & 0.213 & 0.348 & 0.468 & 0.365 & 0.351 & 0.366 & 0.469 \\ 
  SAR & 0.344 & 0.214 & 0.196 & 0.214 & 0.345 & 0.468 & 0.370 & 0.354 & 0.369 & 0.470 \\ 
  GP & 0.347 & 0.217 & 0.199 & 0.217 & 0.354 & 0.470 & 0.372 & 0.356 & 0.373 & 0.476 \\ 
  qgam & 3.985 & 2.461 & 1.877 & 2.269 & 3.062 & 1.256 & 1.266 & 1.195 & 1.266 & 1.425 \\
  \hline
  &\multicolumn{5}{c|}{MCIW}&\multicolumn{5}{c|}{CP}\\
\hline
 & 0.1 & 0.3 & 0.5 & 0.7 & 0.9 & 0.1 & 0.3 & 0.5 & 0.7 & 0.9 \\ 
  \hline
  HS & 1.383 & 1.409 & 1.393 & 1.377 & 1.326 & 0.800 & 0.899 & 0.917 & 0.901 & 0.807 \\ 
  Lap & 1.403 & 1.577 & 1.576 & 1.561 & 1.389 & 0.791 & 0.907 & 0.920 & 0.908 & 0.794 \\ 
  SAR & 1.524 & 1.700 & 1.704 & 1.689 & 1.514 & 0.830 & 0.927 & 0.942 & 0.927 & 0.831 \\ 
  GP & 1.525 & 1.712 & 1.717 & 1.703 & 1.510 & 0.826 & 0.928 & 0.942 & 0.927 & 0.826 \\
\hline
\hline
&\multicolumn{10}{c|}{Scenario (ii)}\\
\hline
&\multicolumn{5}{c|}{MSE}&\multicolumn{5}{c|}{MAD}\\
\hline
 &0.1& 0.3 & 0.5 & 0.7 &0.9&0.1& 0.3 & 0.5 & 0.7 &0.9\\
\hline
HS & {\bf 0.502} & {\bf 0.295} & {\bf 0.264} & {\bf 0.288} & {\bf 0.488} & {\bf 0.494} & {\bf 0.372} & {\bf 0.350} & {\bf 0.369} & {\bf 0.484} \\ 
  Lap & 0.553 & 0.309 & 0.283 & 0.311 & 0.558 & 0.533 & 0.403 & 0.385 & 0.404 & 0.534 \\ 
  SAR & 0.540 & 0.300 & 0.275 & 0.307 & 0.534 & 0.532 & 0.403 & 0.385 & 0.408 & 0.532 \\ 
  GP & 0.545 & 0.310 & 0.282 & 0.309 & 0.549 & 0.533 & 0.409 & 0.389 & 0.410 & 0.538 \\ 
  qgam & 3.973 & 2.419 & 1.891 & 2.219 & 3.208 & 1.315 & 1.241 & 1.197 & 1.256 & 1.459 \\ 
  \hline
  &\multicolumn{5}{c|}{MCIW}&\multicolumn{5}{c|}{CP}\\
\hline
 &0.1& 0.3 & 0.5 & 0.7 &0.9&0.1& 0.3 & 0.5 & 0.7 &0.9\\
\hline
HS & 1.540 & 1.544 & 1.503 & 1.497 & 1.453 & 0.800 & 0.900 & 0.914 & 0.898 & 0.797 \\ 
  Lap & 1.556 & 1.726 & 1.710 & 1.699 & 1.525 & 0.797 & 0.910 & 0.920 & 0.907 & 0.796 \\ 
  SAR & 1.686 & 1.859 & 1.848 & 1.839 & 1.673 & 0.828 & 0.928 & 0.938 & 0.927 & 0.828 \\ 
  GP & 1.681 & 1.864 & 1.858 & 1.852 & 1.662 & 0.823 & 0.924 & 0.937 & 0.924 & 0.823 \\  
  \hline
  \hline
  &\multicolumn{10}{c|}{Scenario (iii)}\\
\hline
&\multicolumn{5}{c|}{MSE}&\multicolumn{5}{c|}{MAD}\\
\hline
 &0.1& 0.3 & 0.5 & 0.7 &0.9&0.1& 0.3 & 0.5 & 0.7 &0.9\\
\hline
HS & {\bf 0.698} & 0.423 & {\bf 0.368} & {\bf 0.391} & {\bf 0.629} & {\bf 0.613} & {\bf 0.473} & {\bf 0.444} & {\bf 0.460} & {\bf 0.585} \\ 
  Lap & 0.765 & 0.434 & 0.387 & 0.422 & 0.755 & 0.659 & 0.498 & 0.473 & 0.494 & 0.652 \\ 
  SAR & 0.747 & {\bf 0.418} & 0.373 & 0.411 & 0.731 & 0.653 & 0.494 & 0.469 & 0.490 & 0.648 \\ 
  GP & 0.760 & 0.422 & 0.378 & 0.416 & 0.768 & 0.656 & 0.497 & 0.472 & 0.494 & 0.661 \\ 
  qgam & 3.792 & 2.295 & 1.908 & 2.174 & 2.984 & 1.366 & 1.258 & 1.204 & 1.255 & 1.422 \\
  \hline
  &\multicolumn{5}{c|}{MCIW}&\multicolumn{5}{c|}{CP}\\
\hline
 &0.1& 0.3 & 0.5 & 0.7 &0.9&0.1& 0.3 & 0.5 & 0.7 &0.9\\
\hline
HS & 1.886 & 1.910 & 1.868 & 1.848 & 1.801 & 0.782 & 0.888 & 0.898 & 0.884 & 0.790 \\ 
  Lap & 1.942 & 2.117 & 2.100 & 2.087 & 1.896 & 0.787 & 0.902 & 0.917 & 0.904 & 0.790 \\ 
  SAR & 2.074 & 2.271 & 2.259 & 2.242 & 2.051 & 0.819 & 0.927 & 0.939 & 0.925 & 0.828 \\ 
  GP & 2.071 & 2.276 & 2.273 & 2.270 & 2.048 & 0.815 & 0.925 & 0.938 & 0.925 & 0.821 \\  
\hline
\end{tabular}
}
\end{center}
\label{table_random_1}
\end{table}

\begin{table}[thbp]
\caption{Average values of MSE, MAD, MCIW, and CP based on $100$ replications for scenarios (iv), (v), and (vi) (exponential function). The minimum values of MSE and MAD are represented in bold. }
\begin{center}
\resizebox{1.0\textwidth}{!}{ 
\begin{tabular}{|c|ccccc|ccccc|}
\hline
&\multicolumn{10}{c|}{Scenario (iv)}\\
\hline
&\multicolumn{5}{c|}{MSE}&\multicolumn{5}{c|}{MAD}\\
\hline
 &0.1& 0.3 & 0.5 & 0.7 &0.9&0.1& 0.3 & 0.5 & 0.7 &0.9\\
\hline 
HS & {\bf 0.183} & 0.125 & 0.114 & 0.119 & {\bf 0.177} & {\bf 0.332} & 0.273 & 0.263 & {\bf 0.267} & {\bf 0.321} \\ 
  Lap & 0.219 & {\bf 0.117} & {\bf 0.106} & {\bf 0.115} & 0.222 & 0.371 & {\bf 0.269} & {\bf 0.257} & 0.268 & 0.370 \\ 
  SAR & 0.237 & 0.123 & 0.109 & 0.118 & 0.226 & 0.391 & 0.278 & 0.263 & 0.274 & 0.382 \\ 
  GP & 0.251 & 0.119 & 0.107 & 0.125 & 0.273 & 0.405 & 0.275 & 0.262 & 0.285 & 0.423 \\ 
  qgam & 0.577 & 0.450 & 0.386 & 0.418 & 0.510 & 0.522 & 0.503 & 0.504 & 0.522 & 0.551 \\ 
  \hline
  &\multicolumn{5}{c|}{MCIW}&\multicolumn{5}{c|}{CP}\\
\hline
 & 0.1 & 0.3 & 0.5 & 0.7 & 0.9 & 0.1 & 0.3 & 0.5 & 0.7 & 0.9 \\ 
  \hline
  HS & 1.092 & 1.097 & 1.095 & 1.098 & 1.107 & 0.788 & 0.873 & 0.882 & 0.876 & 0.815 \\ 
  Lap & 1.216 & 1.235 & 1.229 & 1.232 & 1.213 & 0.832 & 0.924 & 0.939 & 0.928 & 0.842 \\ 
  SAR & 1.342 & 1.381 & 1.368 & 1.366 & 1.330 & 0.869 & 0.951 & 0.960 & 0.952 & 0.878 \\ 
  GP & 1.333 & 1.424 & 1.433 & 1.445 & 1.348 & 0.878 & 0.961 & 0.972 & 0.961 & 0.872 \\
\hline
\hline
&\multicolumn{10}{c|}{Scenario (v)}\\
\hline
&\multicolumn{5}{c|}{MSE}&\multicolumn{5}{c|}{MAD}\\
\hline
 &0.1& 0.3 & 0.5 & 0.7 &0.9&0.1& 0.3 & 0.5 & 0.7 &0.9\\
\hline
HS & {\bf 0.357} & 0.177 & 0.152 & 0.174 & 0.351 & {\bf 0.419} & 0.303 & 0.281 & 0.298 & {\bf 0.405} \\ 
  Lap & 0.384 & 0.169 & 0.139 & 0.161 & 0.373 & 0.446 & 0.300 & {\bf 0.272} & {\bf 0.292} & 0.431 \\ 
  SAR & 0.376 & 0.165 & 0.135 & {\bf 0.155} & {\bf 0.338} & 0.453 & 0.304 & 0.274 & 0.294 & 0.428 \\ 
  GP & 0.385 & {\bf 0.152} & {\bf 0.131} & 0.166 & 0.418 & 0.457 & {\bf 0.293} & {\bf 0.272} & 0.304 & 0.475 \\ 
  qgam & 0.686 & 0.472 & 0.398 & 0.432 & 0.619 & 0.588 & 0.515 & 0.508 & 0.528 & 0.601 \\
  \hline
  &\multicolumn{5}{c|}{MCIW}&\multicolumn{5}{c|}{CP}\\
\hline
 &0.1& 0.3 & 0.5 & 0.7 &0.9&0.1& 0.3 & 0.5 & 0.7 &0.9\\
\hline
HS & 1.249 & 1.192 & 1.189 & 1.203 & 1.242 & 0.777 & 0.878 & 0.898 & 0.880 & 0.798 \\ 
  Lap & 1.392 & 1.344 & 1.315 & 1.318 & 1.350 & 0.824 & 0.923 & 0.941 & 0.927 & 0.829 \\ 
  SAR & 1.541 & 1.495 & 1.458 & 1.454 & 1.491 & 0.856 & 0.945 & 0.959 & 0.949 & 0.866 \\ 
  GP & 1.514 & 1.533 & 1.525 & 1.557 & 1.517 & 0.858 & 0.959 & 0.969 & 0.955 & 0.853 \\  
  \hline
  \hline
  &\multicolumn{10}{c|}{Scenario (vi)}\\
\hline
&\multicolumn{5}{c|}{MSE}&\multicolumn{5}{c|}{MAD}\\
\hline
 &0.1& 0.3 & 0.5 & 0.7 &0.9&0.1& 0.3 & 0.5 & 0.7 &0.9\\
\hline
HS & {\bf 0.450} & 0.242 & 0.201 & {\bf 0.202} & {\bf 0.374} & {\bf 0.492} & 0.371 & 0.345 & {\bf 0.345} & {\bf 0.447} \\ 
  Lap & 0.511 & 0.238 & 0.199 & 0.212 & 0.457 & 0.531 & 0.367 & 0.340 & 0.349 & 0.499 \\ 
  SAR & 0.524 & 0.236 & 0.195 & 0.207 & 0.436 & 0.551 & 0.369 & 0.338 & 0.348 & 0.504 \\ 
  GP & 0.579 & {\bf 0.224} & {\bf 0.182} & 0.221 & 0.587 & 0.578 & {\bf 0.361} & {\bf 0.327} & 0.359 & 0.584 \\ 
  qgam & 0.916 & 0.598 & 0.492 & 0.505 & 0.655 & 0.718 & 0.591 & 0.565 & 0.579 & 0.660 \\ 
  \hline
  &\multicolumn{5}{c|}{MCIW}&\multicolumn{5}{c|}{CP}\\
\hline
 &0.1& 0.3 & 0.5 & 0.7 &0.9&0.1& 0.3 & 0.5 & 0.7 &0.9\\
\hline
HS & 1.502 & 1.441 & 1.411 & 1.409 & 1.497 & 0.767 & 0.855 & 0.876 & 0.872 & 0.813 \\ 
  Lap & 1.696 & 1.630 & 1.585 & 1.578 & 1.646 & 0.812 & 0.910 & 0.929 & 0.923 & 0.842 \\ 
  SAR & 1.884 & 1.842 & 1.776 & 1.748 & 1.804 & 0.852 & 0.947 & 0.959 & 0.949 & 0.876 \\ 
  GP & 1.887 & 1.886 & 1.853 & 1.894 & 1.872 & 0.849 & 0.957 & 0.971 & 0.960 & 0.860 \\  
\hline
\end{tabular}
}
\end{center}
\label{table_random_2}
\end{table}

\section{Additional information for Tokyo crime data analysis}

\begin{figure}[htbp]
\begin{center}
\includegraphics[width=12cm]{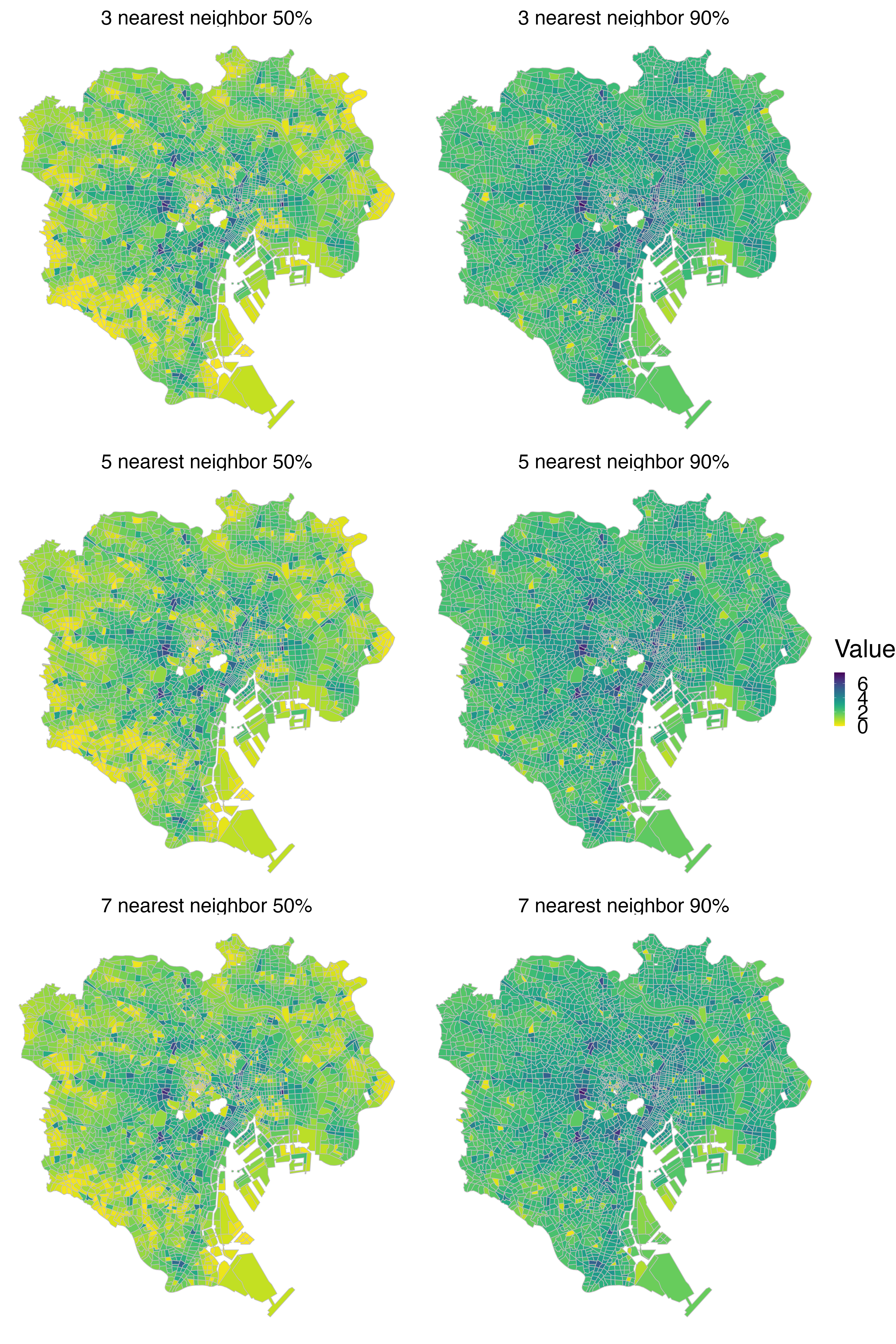}
\caption{The three estimates under 3, 5, and 7 nearest neighbors from top to bottom.}
\label{TokyoCrime_357graph}
\end{center}
\end{figure}

In Section 4 of the main manuscript, the edges were constructed based on the 5 nearest neighbor searches. We compare the results with those of 3 and 7 nearest neighbors. The result is reported in Figure~\ref{TokyoCrime_357graph}. From the figure, we can observe that the results for each number of nearest neighbors do not change very much. The number of edges for each graph is 5598, 8,996, and 12,398. It seems that the graph structure did not affect the smoothness in the example.

\end{document}